\documentclass[11pt]{article}
\usepackage{graphicx}
\usepackage{epsfig}
\usepackage{epsf,amsfonts,amssymb}
\newcommand{\be}{\begin{equation}}
\newcommand{\ee}{\end{equation}}
\newcommand{\bea}{\begin{eqnarray}}
\newcommand{\eea}{\end{eqnarray}}
\newcommand{\ba}{\begin{eqnarray}}
\newcommand{\ea}{\end{eqnarray}}

\newcommand{\beq}{\begin{equation}}
\newcommand{\eeq}{\end{equation}}
\newcommand{\beqa}{\begin{eqnarray}}
\newcommand{\eeqa}{\end{eqnarray}}
\newcommand{\beqar}{\begin{eqnarray*}}
\newcommand{\eeqar}{\end{eqnarray*}}
\newcommand{\lp}{\left(}
\newcommand{\rp}{\right)}
\newcommand{\lpp}{\left[}
\newcommand{\rpp}{\right]}

\newcommand{\eg}{{\it e.g.,}\ }
\newcommand{\ie}{{\it i.e.,}\ }

\def\O{\mathcal{O}}

\begin{document}

\setlength{\unitlength}{1mm}

\thispagestyle{empty}
 \vspace*{1.0cm}

\begin{center}
{\bf \Large Gravitational radiation in $d>4$ from effective field theory}\\

\vspace*{2cm} {\bf Vitor Cardoso,}$^{1\,,\,2}\,$ {\bf \'Oscar
J.~C.~Dias,}$^{3\,,\,4}\,$ {\bf Pau Figueras}$^{\,5}$

\vspace*{0.5cm}

{\it $^1\,$ CENTRA, Dept. de F\'{\i}sica, Instituto
Superior T\'ecnico, \\
Av. Rovisco Pais 1, 1049-001 Lisboa, Portugal}\\[.3em]
{\it $^2\,$Dept. of Physics and Astronomy, The University of
Mississippi, \\
University, MS
38677-1848, USA}\\[.3em]
{\it $^3\,$Departament de F{\'\i}sica Fonamental, Universitat de
Barcelona, \\
Av. Diagonal 647, E-08028 Barcelona, Spain}\\[.3em]
{\it $^4\,$Dept. de F\'{\i}sica e Centro de F\'{\i}sica do Porto,
Faculdade de Ci\^encias da Universidade do Porto, Rua do Campo
Alegre 687, 4169 - 007 Porto, Portugal}\\[.3em]
{\it $^5\,$Center for Particle Theory \& Department of Mathematical Sciences, University of Durham,\\
    Science Laboratories, South Road, Durham DH1 3LE, United Kingdom}\\[.3em]

\vspace*{0.3cm} {\tt vcardoso@fisica.ist.utl.pt, odias@ub.edu,
pau.figueras@durham.ac.uk}

\vspace*{2cm}
 {\bf ABSTRACT}
\end{center}

\vspace{.5cm} Some years ago, a new powerful technique, known as the
Classical Effective Field Theory, was proposed to describe classical
phenomena in gravitational systems. Here we show how this approach
can be useful to investigate theoretically important issues, such as
gravitational radiation in any spacetime dimension. In particular,
we derive for the first time the Einstein-Infeld-Hoffman Lagrangian
and we compute Einstein's quadrupole formula for any number of flat
spacetime dimensions.

\noindent

\vfill \setcounter{page}{0} \setcounter{footnote}{0}
\newpage

\tableofcontents

\setcounter{equation}{0}

\setcounter{equation}{0}\section{\label{sec:intro}Introduction} 
One of the most conceptually simple and interesting problems in any
theory of gravitation concerns the dynamics of two structureless
particles, moving under their mutual gravitational interaction. This
is the two body problem \cite{Blanchet:2001yf}, which has proved to
be fairly simple in Newtonian gravity but a formidable task in
General Relativity. A clear understanding of its solution allows for
tests of Einstein's and alternative theories of gravity
\cite{Will:2001mx,cliffbook}, in the weak and strong field regime.
With the advent of high-sensitivity gravitational wave experiments,
such as LIGO  and others \cite{ligo}, the importance of the two body
problem has exponentially increased. Indeed, one of the main targets
for these experiments is the inspiral and coalescence of neutron
star or black hole binaries. An accurate solution to this two body
problem is necessary not only to compare theory with experiment but
also to improve the chances of detection: the signals expected to
impinge on the detectors will be completely buried in noise, and the
most powerful way to dig them out, matched filtering \cite{matched},
requires a very accurate knowledge of theoretical templates. This
curious interplay between theoretical and experimental needs has
boosted the effort to find accurate solutions to the dynamics of
gravitationally interacting, but otherwise isolated bodies in
General Relativity.

Only recently have long-term stable numerical evolutions of black
hole binaries been achieved \cite{Pretorius:2005gq}, but these still
cover only a very restricted part of the evolution, since they are
extremely demanding from a computational point of view. Even if the
situation improves dramatically in the near future, there's only a
limited amount of computational power and therefore numerical
waveforms always cover only a limited amount of time in the dynamics
of the two body problem. Analytical solutions are required, and they
serve two very important purposes. First of all, analytical
solutions are valuable on their own, as solutions which in some
regime are accurate enough and describe well enough the problem at
hand. They usually provide physical insight that is absent on
numerical calculations. A second important purpose relates to the
validation of numerical codes, allowing simultaneously a extension
of the numerics by some kind of matching procedure. Analytical
solutions to the two body problem have been studied for many decades
now, starting from Einstein himself. The traditionally most powerful
technique, the Post-Newtonian (PN) formalism
\cite{Blanchet:2002av,thorne300,damourbook}, expands the problem as
a series in powers of $v/c$.

In an impressive tour de force, Goldberger and Rothstein
\cite{Goldberger:2004jt,Goldberger:2007hy,Goldberger:2005cd} and
Porto and Rothstein \cite{Porto:2005ac}-\cite{Porto:2007pw} and
later Kol and Smolkin \cite{Kol:2007rx} have shown how to extend and
simplify considerably the PN calculations, using methods borrowed
from the Effective Field Theory (EFT) approach to quantum field
theory. As discussed in \cite{Goldberger:2004jt,Kol:2007rx}, this
EFT of gravity is a \textit{classical} theory, justifying the use of
the acronym ClEFT (Classical Effective Field Theory)
\cite{Kol:2007rx}. Although the computations are carried out in
terms of Feynman diagrams using the quantum field theory language,
no quantum effects are considered. The objects whose physical
properties one is interested in are macroscopic, and hence the
quantum corrections to the classical observables are negligible.
This method is especially well suited to treat the early stages of
the binary inspiral problem when the two objects are
non-relativistic. In this problem there are three widely separated
scales: the typical internal scale of the objects $r_0$, the orbital
distance $r$ between them, and the radiation wavelength $\ell\sim
r/v$, where $v$ is the typical velocity of the system. At the
computational level, having an exact description of the system does
not seem to be practical nor necessary. The reason is that in
gravitational wave physics one is only interested in the radiation
that reaches the detector, and therefore it is sufficient to have an
effective description of the system that captures this information.
This is precisely what the ClEFT does: by systematically integrating
out short distance scales, one is left with an effective description
of the system in terms of the relevant degrees of freedom at the
scale we are interested in. For the binary problem, these are
radiation gravitons coupled to point-particles, and the relevant
length scale is $\ell$. The main advantage of this approach is that
these degrees of freedom are described by a Lagrangian and to obtain
the observables one has to compute Feynman diagrams, as in standard
quantum field theory. Therefore, the procedure is very systematic.

Another advantage of the ClEFT approach is that it has manifest
power counting in the small parameter of the theory, namely,
$v/c$. Therefore, one can systematically compute a given
observable to any desired accuracy since one can determine which
terms  in the Lagrangian and which Feynman diagrams will
contribute to that order. This is important from a practical point
of view because highly accurate theoretical predictions are
required to compare to the putative gravitational wave signals to
be detected by the interferometers. The ClEFT approach to
gravitational radiation also provides a systematic way to deal
with the divergences that plague the PN calculations. These
divergences can be thought of UV divergences that arise because
one is treating extended objects, like black holes or neutron
stars, as being point-like. In EFT these divergences can be simply
dealt with dimensional regularization. Moreover, the finite-size
effects can be easily taken into account by including in the
Lagrangian higher order operators that are consistent with the
symmetries of the problem. The coefficients of these operators are
then determined by a matching calculation in the full theory, see
\cite{Goldberger:2004jt,Goldberger:2005cd} for a more detailed
discussion.

In most realistic situations one expects that the two compact
objects in the binary system are spinning. Therefore, from both a
theoretical and experimental point of view it is important to
incorporate rotation to the ClEFT. This has been done in
\cite{Porto:2005ac}-\cite{Kol:2007bc}. The resulting EFT of gravity
has manifest power counting and one can compute observables using
Feynman diagrams. Therefore, it retains the computational power of
the original ClEFT of gravity of \cite{Goldberger:2004jt}.

Finally we should point out that the ClEFT approach has also been
successfully applied to other problems of interest which involve
widely separated length scales. In particular, it has been shown
that the ClEFT methods are particularly useful to compute, in a
perturbative expansion, the physical parameters of caged black holes
\cite{Kol:2007rx,Chu:2006ce}. In this situation, there are two
length scales: the size of the black hole $r_0$ and the length of
the compact circle $L$, and the small parameter is a power of their
ratio.

What we want to show here is that the ClEFT approach can and
should be used in other contexts as well, in particular in
calculations of gravitational wave emission in higher dimensional
scenarios. Gravitational radiation properties (intensity,
distribution, etc) depend on the dimensionality, curvature, etc.,
of the model under consideration. Thus, the study of gravitational
waves in different higher dimensional scenarios may provide a
means to constraint or distinguish different models. Wave
propagation in higher dimensions is intrinsically different from
the four-dimensional case
\cite{hadamard,SooTie,hassani,Galtsov:2001iv,Mironov:2007nk,Cardoso:2002pa,Cardoso:2003jf}:
waves in flat even-dimensional spacetimes propagate on the light
cone, just as in our usual four-dimensional world, but waves in
odd dimensional spacetime do not. The extension of the notion and
properties of gravitational waves to any even-dimensional
spacetimes is almost straightforward. By isolating the most
important contribution to the Green's function, we recently
computed the field in the wave zone and the quadrupole formula in
general even-dimensional spacetimes \cite{Cardoso:2002pa}. This
formula provides the simplest, lowest-order contribution to
gravitational radiation emission in general situations of physical
interest. It is relevant, for instance, for string-motivated
braneworld scenarios where string-field gravity plays an important
role \cite{branes}.

Unfortunately, wave propagation in general odd-dimensional
spacetimes are much harder to deal with
\cite{hadamard,riesz,copson}. In particular, some of the rules we
are used to are not valid in these spacetimes. For instance, in
odd-dimensional flat spacetimes a propagating wave leaves a
``tail'' behind. This means that a pulse of gravitational waves
(or any other massless field) travels not only along the light
cone but also spreads out behind it, and slowly dies off in tails.
Progress in the study of gravitational waves has been hampered by
this fact.

We will show in this work that the ClEFT approach is able to give us
the correct quadrupole formula in flat, even dimensional spacetimes
and in the same stroke extend that formula to odd-dimensional
spacetimes. The fundamental reason for this, other than the
simplicity of the ClEFT approach is that it works in the momentum
space, where the Green's function have the same functional behavior.
At the same time, we will also derive for the first time the
generalization of the Einstein-Infeld-Hoffman \cite{EIH} Lagrangian
to an arbitrary number of spacetime dimensions.

\vskip0.3cm

The plan of the paper is the following. In section
\ref{sec:procedure} we start by describing the problem and review
the ClEFT approach \cite{Goldberger:2004jt} that we use. In section
\ref{sec:EIH} we find the higher dimensional Einstein-Infeld-Hoffman
Lagrangian. Finally, in section \ref{sec:QF} we determine the
quadrupole formula in a general flat higher dimensional spacetime.
In appendix \ref{sec:FR}, we derive the Feynman rules for the
graviton propagators, graviton vertices and point particle vertices
needed in the main body of the text. In appendix \ref{sec:AUX} we
present some useful relations needed to compute Feynman diagram
contributions. We try to be self-contained in our presentation.

We take $d$ to be the number of spacetime dimensions. Newton's
constant in $d$-dimensions is represented by $G_d$ and is related to
the gravitational coupling $\kappa_g$ through (\ref{Norm:G}). We
take the $(+,-,\cdots,-)$ signature for the Minkowski spacetime. The
speed of light is taken to be $c\equiv 1$.

\setcounter{equation}{0}
 \section{\label{sec:procedure}Description of the problem. Effective field theory approach}

\subsection{Lengthscales in a binary system}

We are interested on finding the first PN correction to the
gravitational interaction between two bodies, and the quadrupole
formula expressing the gravitational energy emitted by a system
moving at low velocities. One of the simplest background to
address simultaneously these issues is a binary system. The
interacting bodies can be black holes or neutron stars with
typical mass $m$. Such system has three lengthscales. To start
with, we have the size $r_0\propto m$ of the gravitational bodies.
Then, we have the typical length separation $r$ between them.
Finally, we have the wavelength $\ell$ of the emitted radiation.
Alternatively, we can make use of the typical velocity $v$ of the
system. Then, its orbital period is $T \sim r/v$, and its
evolution sources the emission of gravitational waves with
frequency $T^{-1}$ and wavelength $\ell\sim r/v$ (here and
henceforth we take unit light velocity, $c\equiv 1$). If the
system's constituents move slowly compared with the speed of
light, $v\ll 1$, we clearly have three widely separated
lengthscales,
\begin{eqnarray}
\label{scales} r_0\ll r\ll \ell \,.
\end{eqnarray}
In this slow motion condition we can study some important properties
of the system using a post-Newtonian (PN) expansion in powers of
$v\ll 1$. Alternatively, as we will do, we can follow the ClEFT
approach, specially tailored for these problems. Both formalisms
consist on systematically solving the Einstein's equations with
non-relativistic (NR) sources by taking a power series expansion in
the small velocity parameter. The three lengthscales are not
independent. Indeed the Virial theorem for a small motion system
governed by an almost Newtonian interaction states that
\begin{eqnarray}
\label{virial} v^2\sim \frac{\kappa_g^2 m}{r^{d-3}} \,,
\end{eqnarray}
where $\kappa_g^2$ is the gravitational coupling proportional to
Newton's constant $G_d$; see (\ref{Norm:G}). This relation between
the kinetic energy and the Newtonian potential will be important
later since it tells us that $\frac{\kappa_g^2 m}{r^{d-3}}$ also
contributes at the same order as $v^2$ in a velocity expansion.

The gravitational field $g_{\mu\nu}^{f}$ created by this system is
a solution of Einstein's equations, which follow from
extremization of the Einstein-Hilbert action
\begin{equation}
S_{EH}[g_{f}] =  \frac{2}{\kappa_g^2}\int d^dx \sqrt{g_f} \,
R[g_{f}],
\end{equation}
where $\sqrt{g_f}$ is the determinant of the metric $g_{\mu\nu}^{f}$
and $R[g_{f}]$ the corresponding Ricci scalar. As is well-known,
this yields a non-trivial non-linear system of differential
equations and no exact solution $g_{\mu\nu}^{f}$ is known for the
two-body problem.

\subsection{Integrate out the internal structure or short lengthscale physics}

In the ClEFT framework we first decompose the full, or exact
gravitational field $g_{\mu\nu}^{f}$ into two components. Take
$g_{\mu\nu}^{s}$ to be the metric that describes the gravitational
field in the short lengthscale region of order $r_0$, and
$g_{\mu\nu}$ the metric encoding a similar information but this
time in the region with lengthscales of order $r$ and larger.
Then, within a good approximation, we can decompose the full
metric as
\begin{equation}
g_{\mu\nu}^{f}=g_{\mu\nu}^{s}+g_{\mu\nu}\,.
\end{equation}
We can now do a first disentanglement of lengthscales, namely we
separate the short lengthscale $g_{\mu\nu}^{s}$ from the long
lengthscale $g_{\mu\nu}$. Within a traditional  effective field
approach, this is achieved by integrating out $g_{\mu\nu}^{s}$
through the path integral
\begin{equation}
\label{Seff1} \exp\lp i S_{eff}[g_{\mu\nu}]/\hbar \rp = \int
  {\cal D}g_{\mu\nu}^{s} \exp\lp i S_{EH}[g^{f}_{\mu\nu},x] /\hbar \rp,
\end{equation}
where $\hbar$ is Planck's constant. In the classical limit,
$\hbar\rightarrow 0$, this path integral reduces in the saddle
point approximation to the computation around the classical
solution. This is the regime of interest for us here, since we
want to use EFT techniques to compute purely classical
(tree-level) results (for a clear discussion of this issue see
\cite{Kol:2007rx}). In this classical limit we can then write
\begin{equation}
\label{Seff:IntStruc} S_{eff}[g,x_a]= S_{EH}[g]+S_{M}[g,x_a],
\end{equation}
where
\begin{eqnarray}\label{EHppActions}
&& S_{EH}[g] =  \int d^dx  \,L[g]
\,,\qquad L[g]=\frac{2}{\kappa_g^2}\,\sqrt{g} \,R[g]\,;\nonumber\\
&& S_{M}[g,x_a]=S_{pp}[g,x_a]+\cdots\,, \qquad
 S_{pp}[g,x_a]=-\sum_a m_a \int d\tau_a \,.
\end{eqnarray}
This effective action describes the motion of the point particle
ensemble in the $g_{\mu\nu}$ background. It preserves the symmetries
of the original theory, namely general coordinate invariance,
worldline reparametrization invariance and, since we do not consider
internal spin degrees of freedom, $SO(d-1)$ invariance. The black
hole or neutron star effective action $S_{M}$ is to leading order
given by the point particle effective action $S_{pp}$, where
$d\tau_a=\sqrt{g_{\mu\nu}(x_a)dx_a^\mu dx^\nu_a}$ is the proper time
along the worldline $x_a^\mu$ of the $a^{\rm th}$ particle. The dots
in $S_{M}$ represent higher order terms that describe non-minimal
couplings of the point particles to the spacetime metric and are
discussed in detail in \cite{Goldberger:2005cd}. They account for
finite size effects. Therefore they are not relevant to compute the
Einstein-Hoffmann-Infeld correction to the Newtonian interaction
neither to determine the quadrupole formula we are interested in.
Indeed, these quantities are clearly independent of the internal or
finite size structure of the gravitational bodies. They would be
essential to address dissipative or absorption processes on the
black hole horizon or star's surface
\cite{Goldberger:2005cd,Porto:2007qi}.

Resuming, after the first disentanglement, where we integrated out
the internal structure of the gravitational bodies, we have a low
energy effective theory of point particles coupled to gravity
described by (\ref{Seff:IntStruc}).

\subsection{Potential and radiation gravitons}
We can now proceed to further disentangle the two other
lengthscales of the problem. Since we are in the regime
(\ref{scales}), \ie $r,\ell\gg r_0$ we can treat the long
lengthscale metric $g_{\mu\nu}$ perturbatively. That is, for
distances much larger than $r_0$ we are in the weak field regime
and we can linearize $g_{\mu\nu}$ around flat spacetime
$\eta_{\mu\nu}$
\begin{eqnarray}\label{Hh:Def}
g_{\mu\nu}&=&\eta_{\mu\nu}+\kappa_g \,\delta g_{\mu\nu}\nonumber\\
 &=&\eta_{\mu\nu}+\kappa_g \lp H_{\mu\nu}+h_{\mu\nu}\rp\,,\qquad
|H_{\mu\nu}|,|h_{\mu\nu}|\ll 1\,.
\end{eqnarray}
We have decomposed the small perturbations $\delta g_{\mu\nu}$ into
two components \cite{Goldberger:2004jt}. The so-called potential
graviton $H_{\mu\nu}$ describes the gravitational field for
lengthscales of order of the orbital length $r$, while $h_{\mu\nu}$
is often denoted as the radiation graviton describing the
gravitational field at larger distances of the order of $\ell\sim
r/v$. Therefore, the potential graviton $H_{\mu\nu}$ is naturally
the mediator of the gravitational interaction between the point
particles, \ie it is responsible for the force that binds the 2-body
system (or $n$-body). On the other hand, the long wavelength
radiation gravitons describe the gravitational waves emitted by the
system during its time evolution. They are the ones that can
propagate to infinity and be detected by an asymptotic observer.

In the regime of (\ref{Hh:Def}), a small $v$ expansion of the
point-particle action $S_{pp}$ yields
\begin{eqnarray}
\label{Spp:exp} S_{pp} &=& \sum_a m_a \int dx^0_a \\
&& \qquad \times  \left[ -\frac{\kappa_g}{2} \delta g_{00} -
\kappa_g \delta g_{0i} {\bf v}_a^i - \frac{\kappa_g}{4} \delta
g_{00} {\bf v}^2_a  - \frac{\kappa_g}{2} \delta g_{ij} {\bf v}_a^i
{\bf v}_a^j +\frac{\kappa_g^2}{8} \delta g_{00}^2 + \frac{1}{2} {\bf
v}^2_a + \frac{1}{8} {\bf v}^4_a +\cdots\right],\nonumber
\end{eqnarray}
where we defined the velocity vector ${\bf
v}_a^i=\frac{dx^i_a}{dx^0_a}$, and we kept only the low order
terms in the expansion. Higher order terms not shown will not be
used. This action describes the non-linear interactions between
the point particles and the gravitational field.

Let's have a closer look at the properties of the two kind of
gravitons. The potential gravitons $H_{\mu\nu}$ have momentum that
scales as $k^{\mu}=(k^0,{\bf k})\sim \lp {v\over r},{1\over
r}\rp$. Indeed, their energy is given by the source's frequency
but, since they have lengthscale $r$, their spatial momentum is
considerably larger, $|{\bf k}|\sim {1\over r}$. Therefore, they
have spacelike momentum, $k^{\mu}k_{\mu}<0$. They are off-shell
and cannot contribute to propagating degrees of freedom to
infinity. As an important consequence, derivatives of potential
gravitons scale as $\partial_\alpha H^{\mu\nu}=\lp
\partial_t H^{\mu\nu}, \partial_i H^{\mu\nu} \rp \sim \lp {v\over
r},{1\over r}\rp H^{\mu\nu}$. That is, in a small velocity expansion
the spatial derivative is one order lower in $v$ than the time
derivative and gives the leading order contribution. To emphasize
this order distinction between time and spatial derivatives it is
useful to work with the Fourier transform of $H_{\mu\nu}$,
\begin{equation}
\label{FourierHk} H_{\mu\nu}(x) = \int_{\bf k} e^{i{\bf k}\cdot {\bf
x}} H_{{\bf k}_{\,\mu\nu}}(x^0)\,,\qquad \int_{\bf k}\equiv\int
{d^{d-1}{\bf k}\over (2\pi)^{d-1}}\,.
\end{equation}
A time derivative acts as $\partial_t H \rightarrow \int \partial_t
H_{\bf k}$ with $\partial_t\sim v/r$, but a spatial derivative gets
now replaced by a spatial momentum factor, $\partial_i H \rightarrow
\int {\bf k}_i H_{\bf k}$ with $|{\bf k}_i|\sim 1/r$. The advantage
of working with the Fourier transform is that the counting of powers
of $v$ is now much more explicit: a term with a time derivative  $
\int \partial_t H_{\bf k}$ is immediately identified to be one order
higher in $v$ than a spatial term $\int {\bf k}_i H_{\bf k}$. This
provides a practical ``visual" advantage when power counting a long
expression.

We now turn to the radiation gravitons. Their momentum scales as
$k^{\mu}=(k^0,{\bf k})\sim \lp {v\over r},{v\over r}\rp$, \ie
since they have lengthscale $r/v$ their spatial momentum has the
same scale as their energy. So, these gravitons are on-shell,
$k^{\mu}k_{\mu}=0$, and propagate at the speed of light to
infinity where they can be detected. Derivatives acting on a
radiation graviton introduce a power of $v/r$, $\partial_\alpha
h_{\mu\nu} \sim {v\over r}h_{\mu\nu}$, \ie time and spatial
derivatives contribute equally in the power expansion of $v$,
contrary to what happens with the potential gravitons. Not less
important, the ClEFT relies on a small velocity expansion. So,
when power counting powers of $v$ we must do a multipole expansion
of the radiation graviton $h_{\mu\nu}(x)$, \footnote{Indices are
lowered with the Minkowski metric $\eta_{\mu\nu}$.}
\begin{eqnarray}
\label{MultExp}
 h_{\mu\nu}(x^0,{\bf x}) =
  h_{\mu\nu}(x^0,{\bf X}) + \delta {\bf x}^i \partial_i
h_{\mu\nu}{\bigl |}_{x=(x^0,{\bf X})}  + {1\over 2}\delta {\bf x}^i
\delta {\bf x}^j
 \partial_i\partial_j h_{\mu\nu}{\bigl |}_{x=(x^0,{\bf X})} + \mathcal{O}(v^3)\,.
\end{eqnarray}
where $\delta {\bf x} = {\bf x} -{\bf X}$ and ${\bf X}=\lp\sum m_a
{\bf x}_a\rp/\sum m_a$ is the center of mass of the system (or any
other reference point of the particle ensemble). That is, we must do
a Taylor expansion of $h_{\mu\nu}(x)$ around the center of mass to
consistently identify the leading and subleading contributions in
the $v$ expansion. Indeed note that $\delta x^i$ scales as $\delta
x^i \sim r$ and the derivatives scale as $\partial_i h_{\mu\nu}\sim
v/r$. Therefore each new term in the multipole expansion contributes
with an extra power of $v$ relative to the previous Taylor term.

Summarizing, the ingredients to an EFT with a well-defined power
counting scheme in powers of $v$ are the decomposition of the
gravitational field into the potential and radiation
contributions, the expansion of the particle action, the Fourier
transform of the potential graviton and finally the multipole
expansion of the radiation graviton.

\subsection{Integrate out the intermediate orbital lengthscale}

To derive an EFT that has manifest velocity power counting rules, we
now need to integrate out the orbital scale, \ie the potential modes
$H_{\mu\nu}$, by computing the functional integral
\begin{eqnarray}
\label{IntOutPot} \exp\lp i S_{NR}[h,x_a]/\hbar\rp  = \int{\cal
D}H_{{\bf_k}_{\mu\nu}} \exp\lp i S_{eff}[g,x_a] /\hbar\rp \,,\qquad
\hbar\rightarrow 0\,,
\end{eqnarray}
where $S_{eff}[g,x_a]$ is defined in (\ref{Seff:IntStruc}) and the
computation is done in the classical limit $\hbar\rightarrow 0$.
In the process, the radiation graviton is treated as a slowly
varying background field. After this operation we are left with a
new effective action $S_{NR}[h,x_a]$ that describes the
non-relativistic (NR) or Post-Newtonian approximation to General
Relativity, in an expansion in powers of $v$. It encodes the
information concerning the gravitational interaction between the
point particles, mediated by the potential gravitons, but also the
coupling between the emitting system and radiation gravitons.

The explicit operation of integrating out the potential gravitons
will be done in the next two sections using the appropriate
Feynman diagrams and techniques. The action $S_{NR}[h,x_a]$, to
the desired order in $v$, is given by a sum over Feynman diagrams
that must satisfy some rules.

\noindent {\bf i)} To start with, note that the worldline
particles will be represented by a solid line. The point particles
worldline's represent very short wavelengths and thus
``infinitely" heavy fields that were integrated out to get
$S_{eff}[g,x_a]$. They thus have no associated propagator
(roughly, their propagator would be $\frac{1}{k^2+m^2}\sim
\frac{1}{m^2}$ since they are heavy fields) and are treated as
background non-dynamical fields. So in the Feynman diagrams we
have no sum over the momentum of the point particles. This
justifies why loops that are closed by the particles are not
quantum loops but give instead a contribution to the tree-level
result. It also justifies that we should only consider diagrams
that keep connected when we remove the particle worldlines.

\noindent {\bf ii)} Next, in the Feynman diagrams, potential
graviton propagators can appear only as internal lines but never
as external lines. The reason being that potential gravitons have
interaction range of the order of only the orbital distance $r$.
Thus, they mediate the gravitational interaction between the
particles but cannot propagate to the asymptotic region.

\noindent {\bf iii)} Finally, the Feynman diagrams can only
contain external radiation graviton lines that propagate to
infinity. These gravitons have long wavelength and do not
contribute to the binding force between particles. So, there are
no internal radiation graviton lines.

Summarizing, integrating out the non-dynamical potential gravitons
generates the gravitational interactions between the NR particles.
In the functional integral (\ref{IntOutPot}), the effective action
for these interactions arise from diagrams with no external
radiation gravitons. This action will be computed in section
\ref{sec:EIH}. To order $\mathcal{O}(v^2)$ it is the
Einstein-Infeld-Hoffmann Lagrangian, $L_{EIH}$.  On the other hand,
the functional integral (\ref{IntOutPot}) also provides the coupling
of radiation gravitons to the particle ensemble through the diagrams
with external radiation gravitons. This action $L_{rad}$ will be
computed in section \ref{sec:Lrad}.

To compute the Feynman diagram contributions we need the Feynman
rules for the graviton propagators, graviton vertices and point
particle vertices. In appendix \ref{sec:FR} we give these rules and
the details of their computation. To find these Feynman rules we
have to introduce the standard gauge fixing and ghost Lagrangian
contributions. We will follow the background field method
\cite{Dewitt,HooftVeltman74} that fixes the gauge in such a way that
preserves the invariance under diffeomorphisms of the background
metric. This gauge fixing scheme guarantees that the obtained action
is gauge invariant. So the gauge fixing action satisfying these
properties, given in Eq. (\ref{Lgf}), must be added to the action in
the rhs of Eq. (\ref{IntOutPot}). There is also a ghost field
contribution, Eq. (\ref{Lgf}), but this only enters in quantum loop
corrections, not considered here. Thus, the ghost contribution will
not be considered in the functional integral (\ref{IntOutPot}).

\subsection{Integrate out the long wavelength radiation scale}

Having obtained the Lagrangian $L_{rad}$ describing the
interaction between radiation and NR particles from $S_{NR}$, we
can finally integrate out the radiation gravitons $h_{\mu\nu}$
through the functional integral
\begin{eqnarray}
\label{IntOutRad} \exp\lp i S_{QF}[x_a]/\hbar\rp = \int{\cal
D}h_{\mu\nu} \exp\lp i S_{rad}[h,x_a]/\hbar\rp\,,\qquad
\hbar\rightarrow 0\,.
\end{eqnarray}
Integrating out $h_{\mu\nu}$ means that we are left with an
effective action $S_{QF}[x_a]$ that depends only on the particle
coordinates. Therefore, this integration arises from Feynman
diagrams that have no external radiation graviton lines. It yields
the desired quadrupole formula. The explicit computation will be
done in section \ref{sec:QFrad}.

\subsection{Power counting scheme\label{sec:pc}}

The graviton field was decomposed into the potential and radiation
components, (\ref{Hh:Def}). Furthermore, we did a small velocity
expansion of the particle action (\ref{Spp:exp}), we took the
Fourier transform (\ref{FourierHk}) of the potential graviton, and
we did a multipole expansion (\ref{MultExp}) of the radiation field.
These operations are key steps in the ClEFT formalism since they
provide the arena for a clear identification of power counting rules
in the small velocity expansion. Having it at hand we can uniquely
assign powers of the velocity parameter $v$ to any Feynman diagram.
We therefore can organize systematically the Feynman diagrams in
powers of $v$ and compute their contribution to our observable up to
the order we wish.  Before computing in next sections the
observables we are interested on, it is useful to list these power
counting rules.

Let us start with the potential graviton. Its propagator, here
represented in short as $\langle H_{{\bf k}_{\,\alpha\beta}}
 H_{{\bf q}_{\,\mu\nu}}\rangle$, is given by (\ref{FR:PgravProp}).
Time scales as $x^0\sim r/v$ and potential graviton momentum goes as
$|{\bf k}|\sim 1/r$. Since a delta function scales as the inverse of
its argument, (\ref{FR:PgravProp}) tells us that the propagator for
$H_{{\bf k}_{\,\mu\nu}}$ goes as
 $\langle H_{{\bf k}_{\,\alpha\beta}} H_{{\bf q}_{\,\mu\nu}}\rangle
 \sim \lpp\delta({\bf k})\rpp \lpp {\bf k}^{-2}\rpp \lpp\delta(x^0)\rpp \sim r^{d-1} r^{2} v/r\sim r^d v$.
Therefore, this fixes the scale of the potential graviton as
$H_{{\bf k}_{\,\mu\nu}}\sim v^{1/2}r^{d/2}$.

Next, consider the radiation graviton whose propagator $\langle
h_{\alpha\beta}h_{\mu\nu}\rangle$ is written in
(\ref{FR:RgravProp}). Since the radiation graviton momentum scales
as $|k^{\mu}|\sim v/r$, this propagator scales as $\langle
h_{\alpha\beta}h_{\mu\nu}\rangle \sim  \lpp d^d k \rpp \lpp
k^{-2}\rpp\sim \lp v/r\rp^{d}\lp v/r\rp^{-2} \sim r^{2-d}v^{d-2}$
and thus the radiation graviton has scale
 $h_{\mu\nu}\sim \lp v/r\rp^{\frac{d-2}{2}}$.

 Finally, to find the scale of $
\kappa_g m$ (where recall that $\kappa_g^2 \propto G_d$) we
introduce the orbital angular momentum,
\begin{eqnarray}
\label{AngMom} L=mvr\,,
\end{eqnarray}
and we use the Virial relation (\ref{virial}) to get $\kappa_g^2
\,m^2 \sim mv^2 r^{d-3}=Lvr^{d-4}$. Thus, $\kappa_g m \sim
L^{1/2}v^{1/2}r^{\frac{d-4}{2}}$. These power counting rules are
summarized in Table \ref{PCtable}.
\begin{table}
\begin{eqnarray}
\nonumber
\begin{array}{||c|c|c|c|c|c||}\hline\hline
 x^0&  \partial_{\mu} & {\bf k} & H_{{\bf k}_{\,\mu\nu}} & h_{\mu\nu}  & \kappa_g m \\
\hline
 r/v & v/r & 1/r      & v^{1/2} r^{d/2} & \lp v/r\rp^{\frac{d-2}{2}} & L^{1/2}v^{1/2}r^{\frac{d-4}{2}}\\
\hline\hline
\end{array}
\end{eqnarray}
\caption{NRGR power counting rules for time coordinate $x^0$,
spacetime derivatives $\partial_{\mu}$, potential spatial momentum
${\bf k}$, potential graviton $H_{{\bf k}_{\,\mu\nu}}$, radiation
graviton $h_{\mu\nu}$ and the coupling $\kappa_g m$. Note that after
the Fourier transform treatment (\ref{FourierHk}) of the potential
gravitons, derivatives of any graviton introduce for sure a factor
of $v/r$. When power counting we should have in mind the useful
relation, $\lpp \kappa_g \rpp \sim \lpp \kappa_g m \rpp
 \frac{vr}{L}$.} \label{PCtable}
\end{table}
Furthermore, in Table \ref{PCtable2} we collect the power counting
rules for the graviton correlation functions and point particle
vertices that are derived and presented in Appendix \ref{sec:FR}.

Summarizing, an inspection of these power counting rules reveals
that tree-level classical results always follow from Feynman
diagrams with a power in the angular momentum of $L^1$ (for the
diagrams contributing to the gravitational interaction between
bodies) and $L^{1/2}$ (for diagrams describing the interaction of
the system with radiation). Quantum loop corrections would
correspond to diagrams with extra powers of $\hbar/L\ll 1$. So, the
orbital angular momentum is the parameter that counts loops in this
classical EFT. Keeping attached to tree-level diagrams we can get
our classical observables up to the desired order in the velocity
$v\ll 1$, having always in mind that loops closed by a ``heavy"
particle worldline are not quantum.

The unique power counting scheme providing a straightforward
systematic analysis, and the clear representation of the computation
provided by the Feynman diagrams are probably the strongest
qualities of the gravitational classical EFT formalism. On the top
of this, the concepts and techniques of regularization and
renormalization are naturally incorporated and can be used to fix
divergencies that appear at the classical level. For example we will
later used dimensional regularization.

\begin{table}
\begin{eqnarray}
\nonumber
\begin{array}{||c|c|c|c|c||}\hline\hline
 \langle h_{\alpha\beta}h_{\mu\nu}\rangle &
 \langle H_{{\bf k}_{\,\alpha\beta}} H_{{\bf q}_{\,\mu\nu}}\rangle &
 \langle H_{{\bf k}_{\,\alpha\beta}} H_{{\bf q}_{\,\mu\nu}}\rangle_{\otimes} &
 \langle H_{{\bf k}_{\,\alpha\beta}} H_{{\bf q}_{\,\mu\nu}} H_{{\bf p}_{\,\gamma\sigma}}\rangle &
 \langle h H_{{\bf k}_{\,\alpha\beta}} H_{{\bf q}_{\,\mu\nu}}\rangle \\
\hline
 r^{2-d}v^{d-2}  & r^d v  & r^d v^3  & r^{3d/2} L^{-1/2} v^{7/2} & r^{d} L^{-1/2}
v^{d+3 \over 2}\\
\hline
\hline\hline
 V_{\mu\nu}^{(1)} H_{\bf k}^{\mu\nu} &
 V_{\mu\nu}^{(2)} H_{\bf k}^{\mu\nu}&
 V_{\mu\nu}^{(3)} H_{\bf k}^{\mu\nu}&
 V_{\mu\nu}^{(4)} H_{\bf k}^{\mu\nu}&
 V_{\alpha\beta\mu\nu}^{(5)} H_{\bf k}^{\alpha\beta}H_{\bf k}^{\mu\nu} \\
\hline
 L^{1/2} v^{0} & L^{1/2} v  &  L^{1/2} v^{2}&  L^{1/2} v^{2} &  v^2\\
\hline\hline
\end{array}
\end{eqnarray}
\caption{Power counting rules for the: radiation graviton propagator
$\langle h_{\alpha\beta}h_{\mu\nu}\rangle$, potential graviton
propagator $\langle H_{{\bf k}_{\,\alpha\beta}}
 H_{{\bf q}_{\,\mu\nu}}\rangle$, $\mathcal{O}(v^2)$ correction to
the potential graviton propagator
 $\langle H_{{\bf k}_{\,\alpha\beta}} H_{{\bf q}_{\,\mu\nu}}\rangle_{\otimes}$,
3-potential graviton correlation function
 $\langle H_{{\bf k}_{\,\alpha\beta}} H_{{\bf q}_{\,\mu\nu}}
 H_{{\bf p}_{\,\gamma\sigma}}\rangle$, 3--radiation-potential graviton
correlation function $\langle h H_{{\bf k}_{\,\alpha\beta}}
 H_{{\bf q}_{\,\mu\nu}}\rangle$, and for the several point-particle vertices
$V_{\mu\nu}^{(i)}$ displayed in (\ref{FR:VertexRules}). These
graviton propagators or correlation functions and point particle
vertices are derived and presented in Appendix \ref{sec:FR}.}
 \label{PCtable2}
\end{table}
%

\setcounter{equation}{0}
 \section{\label{sec:EIH}Einstein-Infeld-Hoffmann Lagrangian in higher dimensions}

In Newton's gravity theory, perturbations propagate at infinite
velocity and only the mass sources the force between gravitational
bodies. Einstein realized that not only the rest energy but also
kinetic and graviton energies contribute to the gravitational
interaction of a system, and that gravitons propagate not
instantaneously but at the speed of light. Newton's gravitational
force has therefore General Relativity corrections that can be
organized systematically in a power expansion in the velocity of the
interacting bodies. The leading $v$-dependent correction of order
$\mathcal{O}(v^2)$ has first computed by Einstein, Infeld and
Hoffmann in 1938 \cite{EIH} (in 4 dimensions) using a PN formalism
and reproduced in \cite{Goldberger:2004jt} and later in
\cite{Kol:2007bc} using the classical EFT approach. In this section
we compute, for the first time, the Einstein-Infeld-Hoffmann
correction to the gravitational interaction between two bodies in
any dimension using the EFT approach introduced in
\cite{Goldberger:2004jt}.

As discussed previously, the potential gravitons govern the
intermediate lengthscale physics at distances of the order of the
orbital length $r$. Therefore they mediate the gravitational
interaction between the NR bodies. The radiation gravitons, being
long wavelength fields when compared to the typical distance between
the particles, do not mediate this interaction. We thus have to
compute the functional integral (\ref{IntOutPot}) keeping only
Feynman diagram contributions with (internal) potential gravitons
and no radiation gravitons.

In subsection \ref{sec:Newton} we first compute the
$\mathcal{O}(v^0)$ Newton interaction and fix our choice for
Newton's constant. Then, in subsection \ref{sec:EIHlagrange} we find
the Einstein-Infeld-Hoffmann correction.

 \subsection{\label{sec:Newton}Newtonian Lagrangian}

\begin{figure}[ht]
\centerline{\includegraphics[width=12cm]{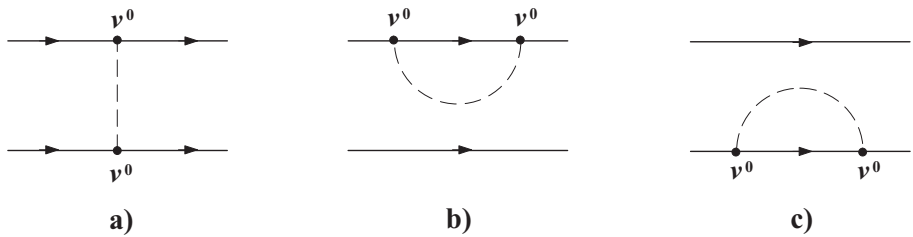}}
\caption{\small Feynman diagrams that give the leading Newtonian
interaction. They are of order $Lv^0$. Actually, the self-energy
diagrams b) and c) give a vanishing contribution after doing
dimensional regularization. Solid lines represent particle
worldlines and dashed lines the potential graviton propagator
(\ref{FR:PgravProp}). Dots are point particle vertices. In these
diagrams they are $V_{\mu\nu}^{(1)}$ in (\ref{FR:VertexRules}). Note
that loops closed by a ``heavy" particle worldline are not quantum
and contribute to tree-level results (\ie we have no integration
over the particle momentum). Quantum loop corrections would
correspond to diagrams with extra powers of $\hbar/L\ll 1$, \ie with
graviton loops. (The reader interested in quantum corrections to the
gravitational interaction between two bodies can see
\cite{Donoghue:1994dn,BjerrumBohr:2002kt,Bohr:2001} and references
therein.) } \label{FigFD:Newton}
\end{figure}

The Feynman diagrams that contribute to leading $\mathcal{O}(Lv^0)$
order are those of Fig. \ref{FigFD:Newton}. The first diagram has a
non-vanishing contribution given by
\begin{eqnarray}
\label{FD:Newton:a} {\rm Fig.~ \ref{FigFD:Newton}a} &=&
V_{(1)}^{\alpha\beta} \langle H_{{\bf k}_{\,\alpha\beta}} (x_2^0)
 H_{{\bf q}_{\,\mu\nu}}(x^0_1)\rangle V_{(1)}^{\mu\nu} \nonumber\\
 &=& {i \kappa_g^2 m_1 m_2 \over 4}\,\frac{d-3}{d-2}\int dx^0_1 dx^0_2\delta(x^0_1-x^0_2)
\int_{\bf k} {1\over {\bf k}^2} e^{-i{\bf k}\cdot\left({\bf x}_1 - {\bf x}_2\right)} \nonumber\\
 &=& i
 \int dt \,\frac{d-3}{d-2}\,
 \frac{\Gamma\lp \frac{d-3}{2}\rp}{16 \,\pi^{\frac{d-1}{2}}} {\kappa_g^2 m_1 m_2\over |{\bf x}_1(t) -
{\bf x}_2(t)|^{d-3}},
\end{eqnarray}
where we used the Feynman rule (\ref{FR:VertexRules}) for the point
particle vertex $V_{(1)}^{\mu\nu}$ and (\ref{FR:PgravProp}) for the
potential graviton propagator $\langle H_{{\bf k}_{\,\alpha\beta}}
H_{{\bf q}_{\,\mu\nu}}\rangle$. We also used relations
(\ref{IntRepdelta})-(\ref{IntMomPos}). In the end we did the
identification $t\equiv x^0_1$. Note that this diagram  scales as $L
v^0$ since $V_{(1)} \langle H_{\bf k}
 H_{\bf q}\rangle V_{(1)}\sim
 \lp L^{1/2} v^{0}\rp^2
\sim L v^0$ (see Table \ref{PCtable2}).

The two last self-energy diagrams in Fig. \ref{FigFD:Newton} are
pure counterterm and have no physical effect. They renormalize the
particle masses and formally vanish after doing dimensional
regularization, as pointed out in \cite{Goldberger:2004jt}. For
example, diagram \ref{FigFD:Newton}b) gives
\begin{equation}
 {\rm Fig.~ \ref{FigFD:Newton}b}=(-1)^d\left(\frac{d-3}{d-2}\right)\,\kappa_g^2\,m_1^2\int dx_1^0 dx_2^0
    \int_\mathbf{k}\,\frac{1}{\mathbf k^2}\;,
\end{equation}
where $\mathbf k$ is the momentum circulating in the loop. The last
integral in the expression above is divergent and it gives rise to a
renormalization of the particle's mass. However, in dimensional
regularization it can be formally set to zero by noting that it
arises as the $\mathbf x\to 0$ limit of the integral
\begin{equation}
 \int \frac{d^{d-1}\mathbf k}{(2\pi)^{d-1}}\, e^{-i\mathbf k\cdot \mathbf x}\,\frac{1}{(\mathbf k^2)^\alpha}=
    \frac{1}{(4\pi)^{\frac{d-1}{2}}}\,
    \frac{\Gamma\left(\textstyle{\frac{d-1}{2}}-\alpha\right)}{\Gamma(\alpha)}\,
    \left(\frac{\mathbf x^2}{4}\right)^{\alpha-\frac{d-1}{2}}\;.
\end{equation}
Sending $\mathbf x\to 0$ before fixing $d$ shows that this integral
vanishes. An obvious equivalent argument also makes
 ${\rm Fig.~\ref{FigFD:Newton}c}=0$.

At this point we specify our normalization for the gravitational
coupling $\kappa_g$ in terms of the $d$-dimensional Newton's
constant $G_d$. We choose, as in $d=4$, to work with the
normalization
\begin{eqnarray}
\kappa_g^2 \equiv 32\pi G_d \,.
 \label{Norm:G}
\end{eqnarray}
Defining the gravitational coupling in this way (\ie in a
$d$-independent form) has the advantage that the black hole entropy
in $d$ dimensions is always horizon area divided by $4G_d$. For this
reason this is the normalization most commonly chosen (see, \eg
\cite{Myers:1986un}, \cite{Cardoso:2002pa},
\cite{Emparan:2008eg})\footnote{Instead, if we prefer to have a
Newtonian potential with $d$-independent pre-factor, $G_d
m/r^{d-3}$, we should have chosen the normalization $\kappa_g^2
\equiv \frac{d-2}{d-3}\,\frac{16 \,\pi^{\frac{d-1}{2}}}{\Gamma\lp
\frac{d-3}{2}\rp}\,G_d$.}. With this choice, the $\mathcal{O}(v^0)$
Newtonian potential energy reads
\begin{eqnarray}
U_N= \frac{d-3}{d-2}\,
 \frac{2\Gamma\lp\frac{d-3}{2}\rp}{\pi^{\frac{d-3}{2}}} {G_d \,m_1 m_2\over |{\bf x}_1(t) -
{\bf x}_2(t)|^{d-3}} \,.
 \label{NewtonPot}
\end{eqnarray}
%

 \subsection{\label{sec:EIHlagrange}Einstein-Infeld-Hoffmann Lagrangian}

\begin{figure}[ht]
\centerline{\includegraphics[width=16cm]{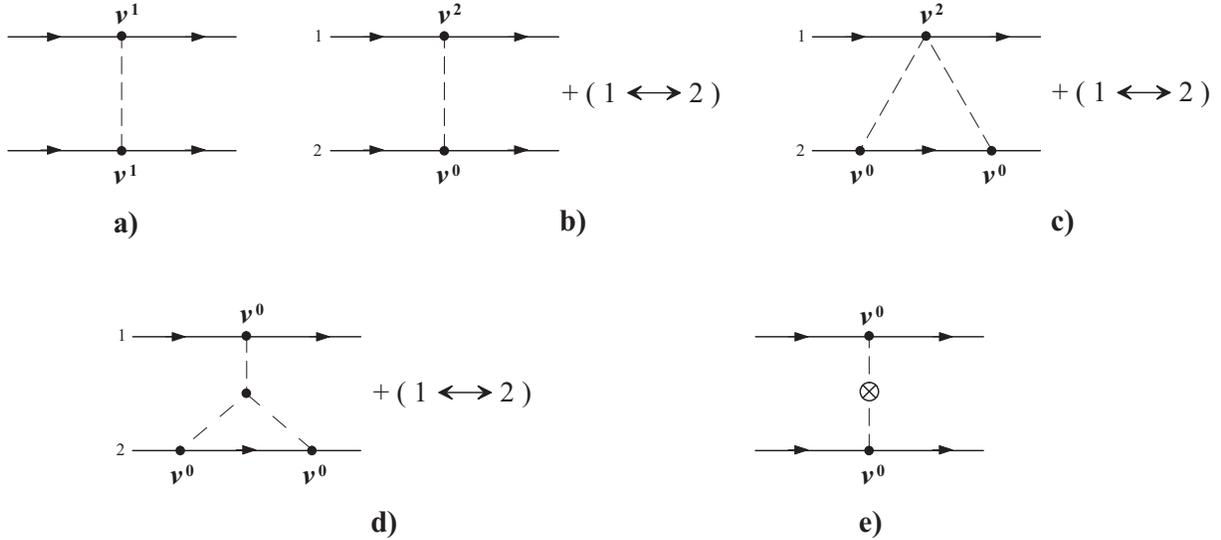}}
\caption{\small  Feynman diagrams for the Einstein-Infeld-Hoffmann
correction to the Newton interaction. They are of order $Lv^2$.
Only (internal) potential gravitons contribute as mediators of the
interaction.} \label{FigFD:EIH}
\end{figure}

The Feynman diagrams contributing to the next-to-leading order
$\mathcal{O}(Lv^2)$ are those of Fig. \ref{FigFD:EIH}. Their sum
gives the Einstein-Infeld-Hoffmann correction to the Newton
interaction.

Using Feynman rules (\ref{FR:VertexRules}) and
(\ref{FR:PgravProp}), respectively for the point particle vertices
$V_{(j)}^{\mu\nu}$ ($j=1,\cdots,5$) and potential graviton
propagator $\langle H_{{\bf k}_{\,\alpha\beta}} H_{{\bf
q}_{\,\mu\nu}}\rangle$, together with relations
(\ref{IntRepdelta}) and (\ref{IntMomPos}) one gets the
contributions from each of the diagrams of  Fig. \ref{FigFD:EIH}.

Starting with the first diagram, one has
\begin{eqnarray}
\label{FD:EIH:a} {\rm Fig.~ \ref{FigFD:EIH}a} &=&
V_{(2)}^{\alpha\beta} \langle H_{{\bf k}_{\,\alpha\beta}} (x_2^0)
 H_{{\bf q}_{\,\mu\nu}}(x^0_1)\rangle V_{(2)}^{\mu\nu} \nonumber\\
 &=& i
 \int dt \:
 \frac{-4\Gamma\lp \frac{d-3}{2}\rp}{\pi^{\frac{d-3}{2}}} {G_d m_1 m_2\over |{\bf x}_1(t) -
{\bf x}_2(t)|^{d-3}}\,\lp {\bf v}_1\cdot {\bf v}_2 \rp \,.
\end{eqnarray}
For the second diagram, there are two point particle vertices,
namely $V_{(3)}^{\mu\nu}$ and $V_{(4)}^{\mu\nu}$ in
(\ref{FR:VertexRules}), scaling as $ L^{1/2}v^{2}$ (see Table
\ref{PCtable2}). Also, there are two mirror contributions resulting
from inter-changing the two particles, $\lp 1 \leftrightarrow 2\rp$.
Taking this into account we get,
\begin{eqnarray}
\label{FD:EIH:b} {\rm Fig.~ \ref{FigFD:EIH}b} &=&
V_{(3+4)}^{\alpha\beta} \langle H_{{\bf k}_{\,\alpha\beta}} (x_2^0)
 H_{{\bf q}_{\,\mu\nu}}(x^0_1)\rangle V_{(1)}^{\mu\nu} +\lp 1 \leftrightarrow 2\rp \nonumber\\
 &=& i
 \int dt \:
 \frac{d-1}{d-2}\frac{\Gamma\lp \frac{d-3}{2}\rp}{\pi^{\frac{d-3}{2}}} {G_d m_1 m_2\over |{\bf x}_1(t) -
{\bf x}_2(t)|^{d-3}}\,({\bf v}^2_1+ {\bf v}^2_2)\,.
\end{eqnarray}
To compute the third diagram we introduce a symmetrization factor of
${1\over 2}$ to account for the symmetry under the exchange of the
two graviton lines, yielding
\begin{eqnarray}
\label{FD:EIH:c} {\rm Fig.~ \ref{FigFD:EIH}c} &=& {1\over 2}
V_{(1)}^{\alpha\beta} \langle H_{{\bf k}_{\,\alpha\beta}} (x_2^0)
 H_{{\bf q}_{\,\gamma\sigma}}(x^0_1)\rangle\,
 V_{(5)}^{\gamma\sigma\lambda\eta} \,\langle H_{{\bf q^{\backprime}}_{\,\lambda\eta}} (x_1^0)
 H_{{\bf p}_{\,\mu\nu}}(x^{\backprime \,0}_2)\rangle
 V_{(1)}^{\mu\nu}+\lp 1 \leftrightarrow 2\rp \nonumber\\
 &=& i
 \int dt \:
 2\lp \frac{d-3}{d-2} \frac{\Gamma\lp \frac{d-3}{2}\rp}{\pi^{\frac{d-3}{2}}} \rp^2 {G_d^2 m_1 m_2(m_1+m_2)\over |{\bf x}_1(t) -
{\bf x}_2(t)|^{2(d-3)}} \,.
\end{eqnarray}
The fourth diagram requires the Feynman rule for the 3-point
correlation function for the interaction between 3-potential
gravitons,  $\langle H_{{\bf k}_{\,00}} H_{{\bf q}_{\,00}} H_{{\bf
p}_{\,00}} \rangle$. This is computed in Appendix
\ref{sec:3PotVertex} and given by (\ref{FR:3grav:Pot2}). The
symmetrization factor of ${1\over 2}$ is also needed. Using Eqs.
(\ref{IntRepdelta}) and (\ref{IntMomPos}) and the integrals
(\ref{IntMom}) we get
\begin{eqnarray}
\label{FD:EIH:d} {\rm Fig.~ \ref{FigFD:EIH}d} &=&  {1\over 2}
V_{(1)}^{\alpha\beta} \langle H_{{\bf k}_{\,\alpha\beta}} (x_1^0)
 H_{{\bf q}_{\,\gamma\sigma}}(x^0_2)H_{{\bf p}_{\,\mu\nu}} (x^{\backprime \,0}_2)\rangle\,
 V_{(1)}^{\gamma\sigma} \,
 V_{(1)}^{\mu\nu}+\lp 1 \leftrightarrow 2\rp \nonumber\\
 &=& -i
 \int dt \,
 \lp\frac{d-3}{d-2}\,
 \frac{2\Gamma\lp \frac{d-3}{2}\rp}
 {\pi^{\frac{d-3}{2}}}\rp^2 \frac{G_d^2 \, m_1 m_2 (m_1+m_2)}
 {|{\bf x}_1 -{\bf x}_2|^{2(d-3)}}\,.
\end{eqnarray}
To get the contribution from diagram of Fig. \ref{FigFD:EIH}e we
need the $v^3$ correction to the potential graviton propagator,
$\langle H_{{\bf k}_{\,\alpha\beta}}
 H_{{\bf q}_{\,\mu\nu}}\rangle_{\otimes}$, given in
 (\ref{FR:PgravPropCorr}). We get
\begin{eqnarray}
\label{FD:EIH:e} {\rm Fig.~ \ref{FigFD:EIH}e}  &=&
V_{(1)}^{\alpha\beta} \langle H_{{\bf k}_{\,\alpha\beta}} (x_2^0)
 H_{{\bf q}_{\,\mu\nu}}(x^0_1)\rangle_{\otimes} V_{(1)}^{\mu\nu}
 \\
 &=& i
 \int dt \:
 \frac{d-3}{d-2}\frac{\Gamma\lp \frac{d-3}{2}\rp}{\pi^{\frac{d-3}{2}}} {G_d m_1 m_2\over |{\bf x}_1(t) -
{\bf x}_2(t)|^{d-3}}\,\lp \,{\bf v}_1\cdot {\bf v}_2  -
(d-3)\frac{({\bf v}_1\cdot {\bf x}_{12} )({\bf v}_2\cdot {\bf
x}_{12})}
 {|{\bf x}_1-{\bf x}_2|^2}  \rp \,,\nonumber
\end{eqnarray}
where  ${\bf x}_{12}\equiv {\bf x}_{1}-{\bf x}_{2}$. To evaluate the
result of acting with the $\partial_0$ operator (appearing in
$\langle H_{{\bf k}_{\,\alpha\beta}}
 H_{{\bf q}_{\,\mu\nu}}\rangle_{\otimes}$) we used the second relation of
(\ref{deltaInt}) together with $\partial_{t_b}{\bf x}_a=\delta^a_b
{\bf v}_a$.

Summing all contributions (\ref{FD:EIH:a})-(\ref{FD:EIH:e}) and
adding also the first relativistic correction to the kinetic
energy (which is also of order $Lv^2$, as can be seen from the
last term in expansion (\ref{Spp:exp})), we finally get
\begin{eqnarray}\label{EIHlagrangian}
&& {\hspace{-0.7cm}}L_{EIH} = {1\over 8}\sum_a m_a {\bf v}^4_a -2
\lp \frac{d-3}{d-2} \frac{\Gamma\lp
\frac{d-3}{2}\rp}{\pi^{\frac{d-3}{2}}} \rp^2 {G_d^2\, m_1
m_2(m_1+m_2)\over |{\bf x}_1(t) -
{\bf x}_2(t)|^{2(d-3)}}     \\
&& +\frac{d-3}{d-2}\,\frac{\Gamma\lp \frac{d-3}{2}\rp}{
\,\pi^{\frac{d-3}{2}}}\,\frac{G_d\, m_1 m_2}{|{\bf x}_1 -{\bf
x}_2|^{d-3}}
 \left[\frac{d-1}{d-3}\,({\bf v}^2_1+ {\bf v}^2_2) -
\frac{3d-5}{d-3}\,{\bf v}_1\cdot {\bf v}_2  - (d-3)\frac{({\bf
v}_1\cdot {\bf x}_{12} )({\bf v}_2\cdot {\bf x}_{12})}
 {|{\bf x}_1-{\bf x}_2|^2} \right]\,.\nonumber
\end{eqnarray}
$L_{EIH}$ is the Einstein-Infeld-Hoffmann correction to the
Newtonian potential in a $d$-dimensional background. For $d=4$,
(\ref{EIHlagrangian}) reduces to the result originally derived in
\cite{EIH} and latter reproduced with the ClEFT in
\cite{Goldberger:2004jt} and \cite{Kol:2007bc}.

A final comment is in order. An improved version of the ClEFT
introduced in \cite{Goldberger:2004jt} was recently proposed
\cite{Kol:2007rx}. This improved version \cite{Kol:2007rx} is
probably the most economic computational way to get the desired
classical observables when the geometry is stationary, by optimizing
the original EFT construction. Its key novelty is the observation
that for stationary geometries one can do a Kaluza-Klein dimensional
reduction along the time direction. Generically and in short, this
introduces a scalar and vector fields that replace the original
tensorial graviton and whose propagators are simpler. This technique
also emphasizes the classical nature of the ClEFT and eliminates
unnecessary quantum features (after all we only want tree-level
results). This method has been used \cite{Kol:2007bc} to recover the
EIH correction in $4$ dimensions. Most of our computations were done
prior to publication of \cite{Kol:2007rx}, which is the reason why
we did not follow this approach here. After the release of the
preprint version of our paper, Kol and Smolkin have derived our
results using their improved ClEFT \footnote{And in the way finding
an incorrect factor in the previous version of our final formula
(\ref{EIHlagrangian}).}. Their results are due to appear in the
latest arXiv version of Ref. \cite{Kol:2007bc}.

\setcounter{equation}{0}
 \section{\label{sec:QF}Quadrupole formula in higher dimensions}

Our next step is to find the Lagrangian $L_{rad}$ describing the
coupling between the NR source and the radiation gravitons, up to
leading order in $v$. As we will soon show, this occurs at order
$Lv^{\frac{d+1}{2}}$ and is done in section \ref{sec:Lrad}. The
quadrupole formula is then computed in section \ref{sec:QFrad}.

 \subsection{\label{sec:Lrad}Lagrangian for the interaction between particles and radiation}

To find $L_{rad}$ we take the small $v$ expansion (\ref{Spp:exp}) of
the point particle NR ensemble in the long wavelength regime, \ie
with $\delta g_{\mu\nu}=h_{\mu\nu}$. In this expansion we just
consider terms that couple to the graviton $h_{\mu\nu}$. These are
the terms that can potentially describe the emission of waves out to
infinity. Finally, we take the multipole expansion (\ref{MultExp})
of the radiation gravitons to organize systematically the Feynman
diagrams in a ladder of powers of $v$.

\begin{figure}[th]
\centerline{\includegraphics[width=12cm]{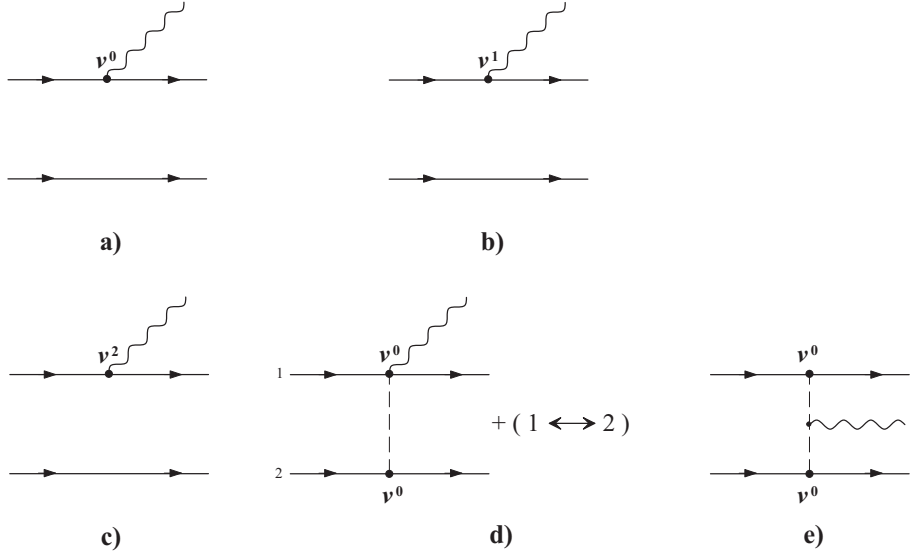}}
\caption{\small Feynman diagrams for the coupling between the NR
source and the radiation gravitons, up to leading order in $v$ where
radiation emission occurs. Diagram a) is of order $\sqrt{L}v^{d-3
\over 2}$, diagram b) is $\mathcal{O}\lp \sqrt{L}v^{d-1 \over
2}\rp$, and c)-e) are order $\sqrt{L}v^{d+1 \over 2}$. Only external
radiation gravitons, represented by wavy lines, describe the
radiated waves.} \label{FigFD:rad}
\end{figure}

Consider the first such term in (\ref{Spp:exp}), describing the
coupling of $h_{00}$ to the mass monopole $\sum m_a$,
\begin{eqnarray}
&& \hspace{-0.5cm}-\sum m_a \!\!\int dx^0_a \frac{\kappa_g}{2} h_{00}(x) \\
&&\qquad =-\sum m_a \!\!\int dx^0_a \frac{\kappa_g}{2} \lp
h_{00}(x^0,{\bf X}) + \delta {\bf x}_a^i
\partial_i h_{00}{\bigl |}_{(x^0,{\bf X})}  + {1\over 2}\delta
{\bf x}_a^i \delta {\bf x}_a^j
 \partial_i\partial_j h_{00}{\bigl |}_{(x^0,{\bf X})} +\mathcal{O}(v^3) \rp
 \,. \nonumber
 \label{Lrad:aAux}
\end{eqnarray}
 Using the power counting rules of Table \ref{PCtable} this
expression scales as $[dx^0][\kappa_g m][h_{00}]\sim \sqrt{L}v^{d-3
\over 2}(1+v+v^2+\cdots)$. We are free to work in the center of mass
(CM) frame where one has ${\bf X}\equiv 0$. Then, the second term in
(\ref{Lrad:aAux}) vanishes, and in the last term one has $\delta
{\bf x}^i ={\bf x}^i-{\bf X}^i={\bf x}^i$. The first term in
(\ref{Lrad:aAux}) gives the leading order contribution to the
coupling between particles and radiation,
\begin{eqnarray}
L_{rad}^{(a)}= -\frac{\kappa_g}{2}
 \sum m_a
h_{00}(x^0,{\bf X})\quad  \rightarrow \quad {\rm Fig.~
\ref{FigFD:rad}a} \,,
 \label{FD:Lrad:a}
\end{eqnarray}
and corresponds to the contribution coming from the Feynman diagram
of Fig. \ref{FigFD:rad}a, that is $\mbox{Fig.
~\ref{FigFD:rad}a}=i\int dt\, L_{rad}^{(a)}$. The third term in
(\ref{Lrad:aAux}) is stored into the $\sqrt{L}v^{d+1 \over 2}$
contribution.

The second coupling term in (\ref{Spp:exp}),
\begin{eqnarray}
&& \hspace{-0.5cm}-\sum m_a \!\!\int dx^0_a \kappa_g
 {\bf v}_a^i h_{0i}(x) \\
&&\qquad =-\sum m_a \!\!\int dx^0_a \kappa_g
 {\bf v}_a^i \lp
h_{0i}(x^0,{\bf X}) + \delta {\bf x}_a^j
\partial_j h_{0i}{\bigl |}_{(x^0,{\bf X})}  +\mathcal{O}(v^2) \rp
 \,, \nonumber
 \label{Lrad:bAux}
\end{eqnarray}
scales as $\sqrt{L}v^{d-1 \over 2}(1+v+\cdots)$. By definition,
the first contribution is proportional to the CM momentum, $\sum
m_a {\bf v}_a^i\equiv {\bf P}_{\rm cm}^i$ and vanishes in the CM
frame, that we choose to work on. Thus,
\begin{eqnarray}
L_{rad}^{(b)}= 0 \quad  \rightarrow \quad {\rm Fig.~
\ref{FigFD:rad}b} \,,
 \label{FD:Lrad:b}
\end{eqnarray}
and the last term in (\ref{Lrad:bAux}) is stored into the
$\sqrt{L}v^{d+1 \over 2}$ contribution. The vanishing of the
$\sqrt{L}v^{d-1 \over 2}$ contribution reflects the absence of
radiation emission of dipole nature in gravity.

The third and fourth terms in (\ref{Spp:exp}) both scale to
leading order as $\sqrt{L}v^{d+1 \over 2}$. This corresponds to
taking only the leading order term $h_{\mu\nu}(x^0,{\bf X})$ in
the multipole expansion of $h_{\mu\nu}(x)$. These two terms
together with the two terms that have been previously stored yield
the following $\sqrt{L}v^{d+1 \over 2}$ contribution to the
interaction Lagrangian between particles and radiation,
\begin{eqnarray}
L_{rad}^{(c)}\!\!&=&\!\! -\frac{\kappa_g}{2}
 \sum m_a \lp {1\over 2}\, {\bf v}^2_a h_{00}(x^0,{\bf X})
 +{\bf v}_a^i {\bf v}_a^j  h_{ij}(t,{\bf X})
 +  {1\over 2}\, {\bf x}_a^i {\bf x}_a^j
 \partial_i\partial_j h_{00}{\bigl |}_{(x^0,{\bf X})}
 + 2{\bf x}_a^i {\bf v}_a^j
\partial_i h_{0j}{\bigl |}_{(x^0,{\bf X})}\rp \nonumber\\
\!\!&\rightarrow&  {\rm Fig.~ \ref{FigFD:rad}c} \,.
 \label{FD:Lrad:c}
\end{eqnarray}
As we shall soon realize, the leading terms in $L_{rad}$
contributing to radiation emission are of order $\sqrt{L}v^{d+1
\over 2}$. Therefore we disregard the other terms in
(\ref{Spp:exp}), all of order higher than $\sqrt{L}v^{d+1 \over
2}$. However, other Feynman diagrams contribute at order
$\sqrt{L}v^{d+1 \over 2}$, namely the diagrams drawn in Fig.
\ref{FigFD:rad}d and Fig. \ref{FigFD:rad}e. They contain both
 internal potential gravitons and external radiation gravitons.

The Feynman diagram \ref{FigFD:rad}d is formally the tensorial
product of diagram \ref{FigFD:Newton}a with diagram \ref{FigFD:rad}a
and yields the contribution \footnote{Using the power counting rules
of Tables \ref{PCtable} and \ref{PCtable2}, diagram \ref{FigFD:rad}d
indeed contributes as
 $ V_{(1)}\langle H_{\bf k} H_{\bf q}\rangle V_{(1)} \kappa_g  h_{00}\sim
\lp  L^{1/2} v^{0}\rp^2  \lp L^{1/2} v^{1/2} r^{\frac{d-4}{2}}
\frac{vr}{L}\rp \lp\frac{v}{r}\rp^{\frac{d-2}{2}} \sim
\sqrt{L}v^{d+1 \over 2}$.}
\begin{eqnarray}
\label{FD:Lrad:d}  {\rm Fig.~ \ref{FigFD:rad}d} &=& \lpp
V_{(1)}^{\alpha\beta} \langle H_{{\bf k}_{\,\alpha\beta}} (x_2^0)
 H_{{\bf q}_{\,\mu\nu}}(x^0_1)\rangle V_{(1)}^{\mu\nu}\rpp
 \lpp -\frac{\kappa_g}{2}h_{00}(x^0,{\bf X})\rpp + \lp 1 \leftrightarrow 2\rp
 \nonumber\\
 &=& i\int dt L_{rad}^{(d)}\quad \rightarrow \quad L_{rad}^{(d)}=-
\frac{d-3}{d-2}\,
 \frac{\Gamma\lp \frac{d-3}{2}\rp}{16 \,\pi^{\frac{d-1}{2}}} {\kappa_g^3 m_1 m_2\over |{\bf x}_1(t) -
{\bf x}_2(t)|^{d-3}}\,h_{00}(x^0,{\bf X}).
\end{eqnarray}

To find the contribution of the diagram shown in Fig.
\ref{FigFD:rad}e we need the 3-point correlation function for the
interaction vertex between two potential gravitons and a radiation
graviton, $\langle h(x^0,X) H_{{\bf k}_{\,00}}(x^0_1) H_{{\bf
q}_{\,00}} (x^0_2)\rangle$. This is computed in section
\ref{sec:FR:vertexPotRad} and given by (\ref{FR:3grav:RadPot2}).
Using also the vertex rule for the coupling
$V_{(1)}^{\alpha\beta}$ between a particle and a potential
graviton we find \footnote{Diagram \ref{FigFD:rad}e scales
as
 $ V_{(1)}\langle h H_{\bf k}
 H_{\bf q}\rangle V_{(1)} \sim
 \lp L^{1/2}v^0/ (r^{d/2} v^{1/2})\rp^2 \lp r^{d} L^{-1/2}
v^{d+3 \over 2} \rp \sim
 \sqrt{L}v^{d+1 \over 2}$, where we used the rules of Table \ref{PCtable2}.}
\begin{eqnarray}
\label{FD:Lrad:e}  {\rm Fig.~ \ref{FigFD:rad}e} &=&
V_{(1)}^{\alpha\beta} \langle h(x^0,X) H_{{\bf k}_{\,\alpha\beta}}
(x_2^0)
 H_{{\bf q}_{\,\mu\nu}}(x^0_1)\rangle V_{(1)}^{\mu\nu}
 \\
&=& i\int dt {\kappa_g^3 m_1 m_2\over 4}\frac{d-3}{d-2}\int_{{\bf
k}} e^{-i{\bf k}\cdot ({\bf x}_1-{\bf x}_2)} {1\over {\bf k}^4} \lpp
{3\over 2}\,{\bf k}^2 h^{00}- \lp {1\over 2}\,{\bf k}^2 \eta_{ij}
+{\bf k}_i{\bf k}_j \rp h^{ij}
 \rpp \nonumber\\
&=& i\int dt\:
 \frac{d-3}{d-2}\,
 \frac{\Gamma\lp \frac{d-3}{2}\rp}{16 \,\pi^{\frac{d-1}{2}}} {\kappa_g^3 m_1 m_2\over |{\bf x}_{12}(t)|^{d-3}}
 \lp {3\over 2} h^{00}
 +{d-3 \over 2}\, \frac{{\bf x}_{12}^i{\bf x}_{12}^j }{|{\bf
x}_{12}(t)|^2} \, h_{ij}\rp = i\int dt L_{rad}^{(e)}
\nonumber\\
\rightarrow   L_{rad}^{(e)}&=& \frac{d-3}{d-2}\,
 \frac{3\Gamma\lp \frac{d-3}{2}\rp}{32 \,\pi^{\frac{d-1}{2}}} {\kappa_g^3 m_1 m_2\over |{\bf x}_1(t) -
{\bf x}_2(t)|^{d-3}}\,h^{00}(x^0,{\bf X})-{\kappa_g \over
2}\sum_{a=1,2} m_a
 {\bf x}_a^i\ddot{{\bf x}}_a^j\,h_{ij}(x^0,{\bf
 X}),\nonumber
\end{eqnarray}
where $\ddot{{\bf x}}\equiv {d^2{\bf x} \over dt^2}$. We used the
momentum integrals (\ref{IntMomPos}), (\ref{IntMomPos2}) and
Newton's laws, ${\bf F}_a=m_a \ddot{{\bf x}}_a$ and ${\bf F}=-\nabla
U_N$, with $U_N$ being Newton's potential energy (\ref{NewtonPot}).
Indeed, they allow to write
\begin{eqnarray}\label{NewtonRel}
\frac{(d-3)^2}{d-2}\,
 \frac{\Gamma\lp \frac{d-3}{2}\rp}{32 \,\pi^{\frac{d-1}{2}}}
  {\kappa_g^3 m_1 m_2\over |{\bf x}_{12}(t)|^{d-3}}
  \frac{{\bf x}_{12}^i{\bf x}_{12}^j } {|{\bf x}_{12}(t)|^2} \, h_{ij}
  \!\!&=&\!\! -{\kappa_g\over 2} \lp {\bf x}_1^i-{\bf x}_2^i\rp
  \lp \nabla^j U_N \rp \,h_{ij}
  \nonumber \\
\!\!&=&\!\!
  -{\kappa_g \over 2}\sum_{a=1,2} m_a
 {\bf x}_a^i\ddot{{\bf x}}_a^j\,h_{ij}\,.
\end{eqnarray}
Summing the contributions from all the Feynman diagrams of Fig.
\ref{FigFD:rad}, \ie adding (\ref{FD:Lrad:a})-(\ref{FD:Lrad:e}) we
get the interaction Lagrangian between particles and radiation up to
order $\sqrt{L}v^{d+1 \over 2}$,
\begin{eqnarray}
S_{rad}=\int dt \, L_{rad}\,, \qquad
L_{rad}=L_{rad}^{(a)}+L_{rad}^{(b)}
+L_{rad}^{(c)}+L_{rad}^{(d)}+L_{rad}^{(e)}+ \cdots \,,
 \label{Lrad:finalAUX1}
\end{eqnarray}
Integrating by parts $L_{rad}^{(d)}$ and the last term of
$L_{rad}^{(e)}$, we can write (\ref{Lrad:finalAUX1}) as
\begin{eqnarray}
L_{rad}\!\!&=&\!\!-{\kappa_g \over 2} {\biggl [}
 \lp \sum_a m_a\lp 1+ {1\over 2}  {\bf v}_a^2\rp- \frac{d-3}{d-2}\,
 \frac{2\Gamma\lp\frac{d-3}{2}\rp}{\pi^{\frac{d-3}{2}}} {G_d \,m_1 m_2\over |{\bf x}_1(t) -
{\bf x}_2(t)|^{d-3}} \rp h^{00} \\
 && + \sum_a m_a\lp  {\bf x}_a^i{\bf v}_a^j-{\bf x}_a^j{\bf v}_a^i\rp \partial_i h_{0j}
 -{1\over 2}\sum_a m_a {\bf x}_a^i{\bf x}_a^j \lp\partial_0\partial_i h_{0j}+\partial_0\partial_j h_{0i}-\partial_0^2h_{ij}
 -\partial_i\partial_j h_{00}\rp {\biggl ]}\,. \nonumber
 \label{Lrad:finalAUX2}
\end{eqnarray}
This Lagrangian can be written in a more suggestive form.
Introducing the orbital angular momentum,
\begin{eqnarray}
{\bf L}_k=\epsilon_{ijk}{\bf x}_a^i{\bf v}_a^j \,,
 \label{Lrad:AngMom}
\end{eqnarray}
one has $\sum m_a\lp  {\bf x}_a^i{\bf v}_a^j-{\bf x}_a^j{\bf
v}_a^i\rp \partial_i h_{0j}= \epsilon^{ijk}{\bf L}_k \partial_j
h_{0i}$,  where we used
$\epsilon^{ijk}\epsilon_{nmk}=\delta^i_{\,m}\delta^j_{\,n}-\delta^i_{\,n}\delta^j_{\,m}$.
To interpret the last term in (\ref{Lrad:finalAUX2}), use the
definition of the quadrupole moment of a gravitational source,
\begin{eqnarray}
I^{ij}=\sum_a m_a {\bf x}_a^i {\bf x}_a^j\,,
 \label{Lrad:momTot}
\end{eqnarray}
and the Riemann tensor $R^{(1)}_{0i0j}$ of the radiated gravitons
(this follows from (\ref{Riemann}), taking Minkowski as the
background metric),
\begin{eqnarray}
R^{(1)}_{0i0j}={\kappa_g \over 2}\lp\partial_0\partial_i
h_{0j}+\partial_0\partial_j h_{0i}-\partial_0^2h_{ij}
 -\partial_i\partial_j h_{00}\rp \,.
 \label{Lrad:Riemann}
\end{eqnarray}
The Lagrangian describing the coupling of radiation to NR sources
can then be written as,
\begin{eqnarray}
L_{rad}=-{\kappa_g \over 2}
 \lpp \sum_a m_a\lp 1+ {1\over 2}  {\bf v}_a^2\rp- U_N \rpp h^{00}
 -{\kappa_g \over 2} \epsilon^{ijk} {\bf L}_k \partial_j
h_{0i} -{1 \over 2}\,I^{ij} R^{(1)}_{0i0j}+\cdots \,,
 \label{Lrad:final}
\end{eqnarray}
where $U_N$ is defined in (\ref{NewtonPot}). The first term in
$L_{rad}$ represents the coupling of $h^{00}$ to the rest mass
plus Newtonian energy of the system. It results from the fact that
in General Relativity, not only the rest mass but also the kinetic
and gravitational potential energy contribute to the effective
mass of the system. This mass correction or monopole term is
conserved at order $\sqrt{L}v^{d-3 \over 2}$. Thus, it does not
contribute to radiation emission, which is a dissipative effect.
The dipole contribution also vanishes as observed in
(\ref{FD:Lrad:b}).

The second term in $L_{rad}$, represents the coupling of the orbital
angular momentum of the system to the radiation gravitons $h_{0i}$,
and is of order $\sqrt{L}v^{d-1 \over 2}$. At this order, the
orbital angular momentum is conserved and thus this term  does not
contribute also to radiation emission.

The leading order term describing radiation emission is the third
term in $L_{rad}$, $-{1 \over 2} I^{ij} R^{(1)}_{0i0j}$, which is of
order $\sqrt{L}v^{d+1 \over 2}$. Note that this term relevant for
the emission process arises from diagram \ref{FigFD:rad}e, which has
3-point graviton self-interaction. This clearly shows that the
quadrupole gravitational emission is sourced by the non-linearities
of General Relativity. Ultimately, this is a consequence of the fact
that graviton energies also source the gravitational field in
General Relativity. In the next subsection we use this radiation
term to find the emitted radiation power. As in the $d=4$ case
\cite{Goldberger:2004jt}, the Lagrangian (\ref{Lrad:final}) is
manifestly gauge invariant under infinitesimal gauge
transformations, showing the power counting scheme in its full
glory.

 \subsection{\label{sec:QFrad}Quadrupole formula}

\begin{figure}[ht]
\centerline{\includegraphics[width=5cm]{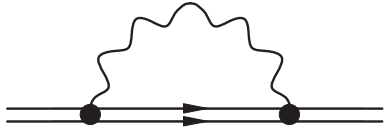}}
\caption{\small Self-energy diagram whose imaginary contribution
gives the quadrupole formula for the radiated power in the form of
gravitational waves by a NR system. The double solid lines represent
the NR particles. In the low energy ClEFT describing the coupling of
the particles to the radiated gravitons, lengthscales smaller than
the orbital distance cannot be resolved. This is pictorially
emphasized in this diagram by drawing the particle lines close to
each other. This diagram is to be understood as a standard
self-energy diagram: at a certain point in the past the system
radiates a graviton (described by (\ref{Lrad:final})) which is then
absorbed back in the future. The amplitude for this process is
complex. Its imaginary part encodes the information on the radiated
energy through gravitational waves that propagates to the asymptotic
region. Again, this is a tree-level diagram since the loop is closed
by ``heavy" particle worldlines whose momenta are not integrated.}
\label{FigFD:SelfE}
\end{figure}

We now want to compute the quadrupole formula that gives the total
energy per unit time radiated through  gravitational waves by a NR
source. The total number of gravitons emitted  by the slowly
motion system over the period $T$ is given by
\begin{eqnarray}
 \int dE d\Omega {d^2\Gamma\over dE d\Omega}={2\over T}
 \,\mbox{Im}\,
S_{QF}[x_a].
\end{eqnarray}
Here, $d\Gamma$ is the differential graviton emission rate, and
$S_{QF}[x_a]$ is the effective action that follows from
integrating out the radiation gravitons through the functional
integral (\ref{IntOutRad}). The imaginary part of this action,
$\mbox{Im}\, S_{QF}$, gives the quantity we are interested in and
is given by the imaginary part of the self-energy Feynman diagram
represented in Fig. \ref{FigFD:SelfE}. Only the last term in
(\ref{Lrad:final}) contributes to radiation emission since this is
the only term with an imaginary part, to leading order. The
self-energy diagram then gives
\begin{eqnarray}\label{FD:selfE1}
\mbox{Fig.}~\ref{FigFD:SelfE} = -{\kappa_g^2 \over 8} \int dx^0_1
dx^0_2 \,I^{ij}(x^0_1) I^{km}(x^0_2) \,\langle
R_{0i0j}^{(1)}(x^0_1,{\bf X}) R_{0k0m}^{(1)}(x^0_2,{\bf X})\rangle.
\end{eqnarray}
%

To find the two-point function of the linearized Riemann tensor we
take the definition (\ref{Lrad:Riemann}) of $R_{0i0j}^{(1)}$ and the
radiation graviton propagator (\ref{FR:RgravProp}), yielding
\begin{eqnarray}\label{2pointF}
\hspace{-0.4cm}\langle R_{0i0j}^{(1)}(x^0_1,{\bf X})
R_{0k0m}^{(1)}(x^0_2,{\bf X})\rangle=\frac{i k_0^4}{k^2+i\epsilon}\,
\frac{d(d-3)}{4(d+1)(d-2)}\lpp {1\over 2}\lp \delta_{ik}\delta_{jm}+
\delta_{im}\delta_{jk} \rp-{1\over
d-1}\,\delta_{ij}\delta_{km}\rpp\!,
\end{eqnarray}
where we used that from the on-shell condition for the radiation
gravitons, $k_0^2-{\bf k}^2=0$, follows that ${\bf k}_i{\bf
k}_j={k_0^2\over d-1}\, \delta_{ij}$ and
 ${\bf k}_i{\bf k}_j{\bf k}_k{\bf k}_m={k_0^4\over d^2-1}
 \lp
 \delta_{ij}\delta_{km}+\delta_{ik}\delta_{jm}+\delta_{im}\delta_{jk}\rp$.
Note that the two-point function in (\ref{2pointF}) is proportional
to the projection operator onto symmetric and traceless two-index
spatial tensors. So when it acts on the quadrupole moment in
(\ref{FD:selfE1}), only the traceless part of the quadrupole moment
$I^{ij}$ will effectively contribute. This traceless component is
given by
\begin{eqnarray}
Q^{ij}=\sum_a m_a \lp {\bf x}_a^i {\bf x}_a^j -{1 \over d-1}\,{\bf
x}_a^2 \delta^{ij}\rp .
 \label{Lrad:Qij}
\end{eqnarray}
Introducing its Fourier transform,
\begin{eqnarray}
Q^{ij}(k_0) = \int dx^0 Q^{ij}(x^0) e^{-i k_0 x^0}\,,
\end{eqnarray}
and using (\ref{2pointF}), (\ref{FD:selfE1}) becomes
\begin{eqnarray}\label{FD:selfE2}
\mbox{Fig.}~\ref{FigFD:SelfE} = -i\,{\kappa_g^2 \over 32}
 \,
 \frac{d(d-3)}{(d+1)(d-2)}\int\frac{d^dk}{(2\pi)^d}\,\frac{(k^0)^4}{k^2+i\epsilon}
 \left| Q_{ij}(k_0)\right|^2,
\end{eqnarray}
with $\left| Q_{ij}\right|^2=Q^{ij}Q^*_{ij}$. To find the
imaginary part of (\ref{FD:selfE2}), that gives $\mbox{Im}\,
S_{QF}$, one uses the well-known relation for the principal value
of a function,
\begin{eqnarray}\label{Princvalue}
&& \int dk^0\,\frac{f(k^0)}{k_0^2-{\bf k}^2+i\epsilon}=P\int
dk^0\,\frac{f(k^0)}{k_0^2-{\bf k}^2}\mp i \pi \int
dk^0\,f(k^0)\,\delta\lp k_0^2-{\bf
k}^2\rp,\nonumber\\
 && \delta\lp k_0^2-{\bf k}^2\rp= {1\over 2|k^0|}\,
 \delta\lp k^0-|{\bf k}|\rp\delta\lp k^0+|{\bf k}|\rp,
\end{eqnarray}
where $P$ stands for the the principle value. This relation extracts
the imaginary contribution of diagram \ref{FigFD:SelfE},
\begin{equation}
\mbox{Im}\, S_{QF}  = -{\kappa_g^2 \over 64} \,
 \frac{d(d-3)}{(d+1)(d-2)}\int\frac{d^{d-1}k}{(2\pi)^{d-1}}\, |{\bf k}|^3 |Q_{ij}(|{\bf k}|)|^2.
\end{equation}
Here, the integrand gives the differential graviton emission rate
along the evolution of the system. The differential power radiated
is then this emission rate times the energy $|{\bf k}|$ of the
graviton. And the total power radiated during the period
$T\rightarrow \infty$ is then
\begin{eqnarray}
P =\int dE d\Omega\, E\,{d^2\Gamma\over dE d\Omega}={2E\over T}
\,\mbox{Im}\, S_{QF}[x_a],
\end{eqnarray}
that is
\begin{eqnarray}
P =-{\kappa_g^2 \over 32 T} \,
 \frac{d(d-3)}{(d+1)(d-2)}\int\frac{d^{d-1}k}{(2\pi)^{d-1}}\, |{\bf k}|^4 |Q_{ij}(|{\bf k}|)|^2.
\end{eqnarray}
Exploring the spherical symmetry of the problem one has $\int
d^{d-1}{\bf k}=\Omega_{d-2}\int dk |{\bf k}|^{d-2}$, with
$\Omega_{d-2}=\frac{2\pi^{d-1\over 2}}{\Gamma\lp d-1\over 2\rp}$
being the area of the unit $(d-2)$-sphere. Moreover, the on-shell
condition for the graviton momentum allows one to make the
replacement $|{\bf k}|=-k^0$. Using $k^0\equiv \omega$, and the
normalization (\ref{Norm:G}) for the gravitational coupling, the
energy radiated per unit of frequency finally reads
\begin{eqnarray} \label{QF:end}
\frac{dE}{d\omega}
=G_d~\frac{2^{2-d}\pi^{-(d-5)/2}d(d-3)}{(d-2)(d+1)\Gamma\left[\textstyle\frac{d-1}{2}\right]}~
\omega^{d+2}|Q_{ij}(\omega)|^2.
\end{eqnarray}
This is the celebrated quadrupole formula, valid for any spacetime
dimension $d>3$. For $d=4$ it was originally computed by Einstein
and reproduced using the EFT formalism in \cite{Goldberger:2004jt}
and \cite{Kol:2007bc}. For even $d$, our formula agrees with the
expression first computed in \cite{Cardoso:2002pa} using a standard
Green's function formalism. Using the ClEFT approach we have shown,
for the first time, that the formula is also valid for odd
$d$-dimensional spacetimes. This was an open problem up to now
\cite{Cardoso:2002pa} and this is thus one of our main results.

\vskip 0.3cm \noindent {\bf Relation with previous work} \vskip
0.2cm

The reason for the difficulties to derive the quadrupole formula in
the traditional approach is the following
\cite{Cardoso:2002pa,Cardoso:2003jf}. Wave propagation of massless
fields is considerably different depending on whether the number of
spacetime dimensions $d$ is even or odd. This follows from the fact
that Green's function has support only on the light-cone in the
former case, but also inside of it in the latter background. Indeed,
in even $d$ the Green's function is roughly \cite{Cardoso:2002pa}
$G^{\rm ret}\sim
\partial_r^{(d-4)/2} \lpp r^{-1}\delta(t-r)\rpp$ (where $r$ stands
for radial distance and $t$ for time), while in odd $d$ one has
$G^{\rm ret}\sim
\partial_r^{(d-3)/2} \lpp \frac{\Theta(t-r)}{\sqrt{t^2-r^2}}\rpp$
(where $\Theta(x)$ is the step function). As a consequence, wave
tails or wakes develop in odd dimensions \cite{Cardoso:2003jf}.
Now, traditional methods to compute the quadrupole formula require
solving the linearized Einstein field equations in the harmonic
gauge, $\square h_{\mu\nu}=16\pi G_d S_{\mu\nu}$, where $\square$
is the $d$-dimensional Laplacian, and $S_{\mu\nu}=
T_{\mu\nu}-\frac{1}{d-2}\,\eta_{\mu\nu}\,T$ encodes the
information on the energy-momentum tensor of the NR source. The
general solution of this inhomogeneous differential equation  can
be expressed with the help of the retarded Green's function
$G^{\rm ret}(t-t', {\bf x - x'})$, as
\begin{equation}
h_{\mu\nu}(t,{\bf x})= 16 \pi G_d \int dt'\int d^{d-1}{\bf x'}
S_{\mu \nu}(t',{\bf x'}) G^{\rm ret}(t-t', {\bf x - x'}) \,,
\label{retsol}
\end{equation}
plus any solution to the homogeneous equation. For even $d$
dimensions, we have a trivial integral over a delta function.
However, in odd $d$, the structure of the Green's function
displayed above makes the integral in (\ref{retsol}) highly
non-trivial, and so far this integration has not performed.

In a forthcoming publication \cite{CDF} we will show that the
quadrupole formula (\ref{QF:end}) can also be reproduced using a
more traditional Green's function formalism. The key observation
will be that the Green's function in {\it momentum} space has the
same functional form for both even and odd spacetimes. So, carrying
the calculations in momentum space instead of in the position space
(as is standard) allows to re-derive (\ref{QF:end}). We will further
discuss in detail the differences in even and odd $d$ spacetimes,
and explore (\ref{QF:end}).


\section*{Acknowledgments}

\noindent We warmly thank Barak Kol and Roberto Emparan for a
careful reading of the manuscript and for very useful comments and
suggestions. We also thank Marco Caldarelli, Henriete Elvang, Mukund
Rangamani, and Joan Soto for useful correspondence. We are grateful
to Niels Bjerrum-Bohr for kindly providing reference
\cite{Bohr:2001}. OJCD thanks the Niels Bohr Institute for
hospitality and the organizers of the workshop ``Mathematical
Aspects of General Relativity", Copenhagen, April 2008, and
``Quantum Black Holes, Braneworlds and Holography", Valencia, May
2008, where part of this work was done. This work was partially
funded by Funda\c c\~ao para a Ci\^encia e Tecnologia (FCT) -
Portugal through projects PTDC/FIS/64175/2006 and
POCI/FP/81915/2007. VC is partially funded by a Fulbright
Scholarship. OJCD acknowledges financial support provided by the
European Community through the Intra-European Marie Curie contract
MEIF-CT-2006-038924.


\appendix

\section*{Appendices}

\setcounter{equation}{0}
 \section{Feynman rules for propagators and vertices}
 \label{sec:FR}

To compute the Feynman diagram contributions of Figs.
\ref{FigFD:Newton}, \ref{FigFD:EIH} and \ref{FigFD:rad} we need
Feynman rules for the graviton propagators, graviton vertices and
point particle vertices. In this appendix we give these rules and
details of their computation. The derivation is self-contained. We
adopt the background field method introduced by Dewitt
\cite{Dewitt} in 1967 and fully developed by t'Hooft and Veltman
\cite{HooftVeltman74}.

We start by considering metric perturbations $\delta g_{\mu\nu}$
around the unperturbed background $g^{(0)}_{\mu\nu}$, up to the
order we will be interested,\footnote{We use the $(+,-,\cdots,-)$
signature so $\sqrt{g}=\sqrt{\pm {\rm det}\,g_{\mu\nu}}$ for odd and
even $d$, respectively.}
\begin{eqnarray}
\label{metricpert}
 g_{\mu\nu}&=&g^{(0)}_{\mu\nu}+\kappa_g \delta g_{\mu\nu}\,, \qquad   g^{\mu\nu}=g_{(0)}^{\mu\nu}-\kappa_g
\delta g^{\mu\nu}+\kappa_g^2 \delta g^{\mu}_{\:\:\alpha}\delta
g^{\alpha\nu}+\O{(\delta g^3)},
\nonumber\\
 \sqrt{g}&=&\sqrt{g_{(0)}}\,e^{\frac{1}{2}{\rm Tr}[\,\ln(\delta^{\mu}_{\:\:\nu}+\kappa_g
 \delta g^{\mu}_{\:\:\nu})\,]}\nonumber\\
 &=&\sqrt{g_{(0)}}  \left[1+\frac{\kappa_g}{2}\,
 \delta g-\frac{\kappa_g^2}{4}\left(\delta g_{\alpha\beta}\delta g^{\alpha\beta}-\frac{1}{2}\delta g^2\right)\right]+\O{(\delta g^3)}\,,
\end{eqnarray}
where we used $g^{\mu\gamma}g_{\gamma\nu}=\delta^{\mu}_{\:\:\nu}$,
$\delta g=g^{\mu\nu}_{(0)}\delta g_{\mu\nu}$, ${\rm det}\,M= e^{{\rm
Tr}[\,\ln M]}$, and the Taylor expansion for $\ln(1+x)$ and $e^x$.

The metric perturbation (\ref{metricpert}) naturally induces
perturbations on the affine connections, Riemann and Ricci tensors.
The perturbations introduced in these tensors can be found in
\cite{HooftVeltman74} and a nice reference to find the details of
their computation is \cite{Bohr:2001}. Here we present the first
order perturbation in the Riemann tensor, because we need this
quantity to get (\ref{Lrad:Riemann}),
\begin{eqnarray}
&& \hspace{-0.6cm}
 ^{(1)}R^{\mu}_{\:\nu\alpha\beta}={\kappa_g \over 2}
 \lp\nabla_\alpha
\nabla_\nu \delta g^\mu_{\:\beta}- \nabla_\beta \nabla_\nu \delta
g^\mu_{\:\alpha}-\nabla_\alpha \nabla^\mu \delta
g_{\nu\beta}+\nabla_\beta \nabla^\mu \delta g_{\nu\alpha} +
 ^{(0)}\!\!R^{\mu}_{\:\gamma\alpha\beta}\,\delta g^\gamma_{\:\nu}
 + ^{(0)}\!\!R^{\gamma}_{\:\nu\beta\alpha}\,\delta
 g^\mu_{\:\gamma}\rp. \nonumber\\
 &&
 \label{Riemann}
\end{eqnarray}
It also induces perturbations on the  Ricci scalar. Expanding the
Einstein-Hilbert Lagrangian (\ref{EHppActions}) in powers of the
gravitational field $\delta g_{\mu\nu}$ one finds that $L[g]=L^{(0)}
+L^{(1)}+L^{(2)}+\mathcal{O}(\delta g^3)$ with:
$L^{(0)}=2R^{(0)}/\kappa_g^2$, $L^{(1)}=0$ (due to the unperturbed
Einstein's equations), and
\begin{eqnarray}
\label{L(2)} L^{(2)}&=&\sqrt{g_{(0)}}{\biggl[}
\frac{1}{2}\left(\frac{1}{2}\delta g^2-\delta
g^{\mu}_{\:\:\nu}\delta g^{\nu}_{\:\:\mu}\right)R^{(0)}
+\left(2\delta g^{\mu}_{\:\:\alpha}\delta g^{\alpha\nu}-\delta g\,
\delta g^{\mu\nu}\right)R^{(0)}_{\mu\nu} -\frac{1}{2}\left(
\nabla_{\alpha}\delta g\right)\left(\nabla^{\alpha}\delta g\right)   \nonumber\\
&& \qquad\quad +(\nabla_{\mu}\delta g)\left( \nabla_{\nu}\delta
g^{\mu\nu}\right) +\frac{1}{2}\left( \nabla_{\alpha}\delta
g_{\mu\nu}\right)\left( \nabla^{\alpha}\delta
g^{\mu\nu}\right)-\left( \nabla_{\alpha}\delta
g^{\mu\nu}\right)\left( \nabla_{\mu}\delta
g^{\alpha}_{\:\:\nu}\right) {\biggr]} \,,
\end{eqnarray}
after partial integrations (PI) and moduli total divergencies (MTD)
in the action integral.

At this point we have to choose the gauge. The background field
method adopts the viewpoint in which the background unperturbed
metric $g_{(0)}^{\mu\nu}$ is left invariant under difeomorphism
transformations, $x^{\mu}\rightarrow x^{\mu}-\xi^{\mu}$ (where
$\xi^{\mu}$ infinitesimal vector field), while the metric
perturbations transform as $\delta g_{\mu\nu}\rightarrow \delta
g_{\mu\nu}+ \nabla_{\mu}\xi_{\nu}+ \nabla_{\nu}\xi_{\mu}$ (where
$\nabla_{\mu}$ is the covariant derivative w.r.t.
$g^{(0)}_{\mu\nu}$). Hence its name. Choosing to work in the
harmonic gauge, $\nabla_{\alpha}\delta
g^{\alpha}_{\:\:\mu}-\frac{1}{2}\nabla_{\mu}\delta g=0$, one then
has the gauge fixing Lagrangian and associated ghost Lagrangian
\cite{HooftVeltman74},
\begin{eqnarray}
\label{Lgf} &&L_{\rm GF}=\sqrt{g_{(0)}}\left(\nabla_{\alpha}\delta
g^{\alpha}_{\:\:\mu}-\frac{1}{2}\nabla_{\mu}\delta g\right)
\left(\nabla_{\alpha}\delta g^{\alpha\mu}-\frac{1}{2}\nabla^{\mu}\delta g\right),    \nonumber\\
&& L_{\rm ghost}= \sqrt{\eta}\,
\eta^{*\,\mu}\left(\nabla_{\alpha}\nabla^{\alpha}\eta_{\mu\nu}-R^{(0)}_{\mu\nu}\right)\eta^{\nu}\,,
\end{eqnarray}
where $\eta^{\mu}$ is the fermionic ghost field. The ghost
contribution would be important only if we were interested in
computing quantum loop corrections. Since we only want tree-level
results we will make no further reference to it.

Defining $L_{\delta g^2}\equiv L^{(2)} + L_{\rm GF}$ one then has
\begin{eqnarray}
\label{LgfLghost} L_{\delta g^2} &=&\sqrt{g_{(0)}}{\biggl[}
\frac{1}{2}\left(\frac{1}{2}\delta g^2-\delta
g^{\mu}_{\:\:\nu}\delta g^{\nu}_{\:\:\mu}\right)R^{(0)}
+\left(2\delta g^{\mu}_{\:\:\alpha}\delta g^{\alpha\nu}-\delta g\,
\delta g^{\mu\nu}\right)R^{(0)}_{\mu\nu}
 +\frac{1}{2}\left(
    \nabla_{\alpha}\delta g_{\mu\nu}\right)\left(\nabla^{\alpha}\delta g^{\mu\nu}\right)  \nonumber \\
&&
 \qquad\quad -\left( \nabla_{\alpha}\delta g^{\mu\nu}\right)\left(
\nabla_{\mu}\delta g^{\alpha}_{\:\:\nu}\right)
 -\frac{1}{4}\left(\nabla_{\alpha}\delta g\right)\left(\nabla^{\alpha}\delta g\right)
 + \left( \nabla_{\alpha}\delta g^{\alpha}_{\:\:\mu}\right)\left(
\nabla_{\nu}\delta g^{\nu\mu}\right) {\biggr]} \,,
\end{eqnarray}
where the covariant derivatives $\nabla_{\mu}$ are taken w.r.t the
background unperturbed metric $g_{(0)}^{\mu\nu}$.

At this point, as motivated in discussion before (\ref{Hh:Def}), we
take the unperturbed background to be Minkowski spacetime,
$g^{(0)}_{\mu\nu}\equiv\eta_{\mu\nu}$. After PI and MTD we then have
\begin{eqnarray}
\label{Lmink} L_{\delta g^2}
=\frac{1}{2}\left(\partial_{\alpha}\delta
g_{\mu\nu}\right)\partial^{\alpha}\delta g^{\mu\nu}
 -\frac{1}{4}\left(\partial_{\alpha}\delta g\right)\partial^{\alpha}\delta g\,.
\end{eqnarray}
%

\subsection{Radiation graviton propagator}

So far, our discussion is independent on the lengthscale of the
gravitational perturbation. We now take the graviton perturbation
to be a long wavelength radiation graviton, \ie we make the
replacement $\delta g_{\mu\nu} \rightarrow h_{\mu\nu}$. Then,
again after PI and MTD, (\ref{Lmink}) reads
\begin{eqnarray}
\label{Lend} L_{h^2}
=-\frac{1}{2}h_{\alpha\beta}\mathcal{D}_h^{\alpha\beta\mu\nu}h_{\mu\nu}\,,
\qquad {\rm with}\quad
 \mathcal{D}_h^{\alpha\beta\mu\nu}=
 \frac{1}{2}\left( \eta^{\alpha\mu}\eta^{\beta\nu}
 +\eta^{\alpha\nu}\eta^{\beta\mu}
 -\eta^{\alpha\beta}\eta^{\mu\nu}\right)\partial_{\lambda}\partial^{\lambda}\,,
\end{eqnarray}
and the associated action is $S_{h^2}=\int d^dx L_{h^2}$.

To find the graviton propagator we have to get the symmetric inverse
of the bilinear operator $\mathcal{D}_h^{\alpha\beta\mu\nu}$. This
amounts to find the inverse tensor $P_{\gamma\sigma\alpha\beta}$
that multiplied by the tensorial part of
$\mathcal{D}_h^{\alpha\beta\mu\nu}$ gives the symmetric identity
$I_{\gamma\sigma}^{~~~\mu\nu}$ defined in (\ref{IdDef}), and to find
the inverse of the differential operator
$\partial_{\lambda}\partial^{\lambda}$. This is simply the Feynman
propagator for a massless boson, $D_F(x-y)$. The Feynman rule for
the radiation graviton propagator is then
%
\begin{eqnarray}
\hspace{-2cm}
 \vtop{  \vskip-11pt  \hbox{\includegraphics[width=3cm]{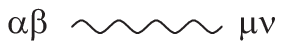} } }
 \:\: &=& \:\:
 \langle h_{\alpha\beta}(x)h_{\mu\nu}(y) \rangle \nonumber\\
&=& D_F(x-y)\, P_{\alpha\beta\mu\nu}\,,
 \label{FR:RgravProp}
\end{eqnarray}
with
\begin{eqnarray}
&&D_F(x-y)=\int\frac{d^dk}{(2\pi)^d}\,\frac{i}{k^2+i\epsilon}\,e^{-i
k \cdot (x-y)}\,,\nonumber\\
 &&P_{\alpha\beta\mu\nu}= \frac{1}{2}\left(
\eta_{\alpha\mu}\eta_{\beta\nu}
 +\eta_{\alpha\nu}\eta_{\beta\mu}
 -\frac{2}{d-2}\,
 \eta_{\alpha\beta}\eta_{\mu\nu}\right)\,.
 \label{FR:RgravPropAUX}
\end{eqnarray}
This propagator has power counting scale $\langle
h_{\alpha\beta}h_{\mu\nu}\rangle \sim r^{2-d}v^{d-2}$, as shown in
section \ref{sec:pc}.

\subsection{Potential graviton propagator. Correction to the potential propagator}

To find the potential graviton propagator we go back to
(\ref{Lmink}). This time we take the graviton perturbation to be a
potential graviton, \ie we do the replacement $\delta g_{\mu\nu}
\rightarrow H_{\mu\nu}$,
\begin{eqnarray}
\label{Lmink:H} L_{H^2}
=\frac{1}{2}\left(\partial_{\alpha}H_{\mu\nu}\right)\partial^{\alpha}H^{\mu\nu}
 -\frac{1}{4}\left(\partial_{\alpha}H\right)\partial^{\alpha}H\,.
\end{eqnarray}
Introducing the Fourier transform (\ref{FourierHk}) of $H_{\mu\nu}$,
and using the integral representation of the delta function
(\ref{IntRepdelta}) one gets
\begin{eqnarray}\label{LpotAUX}
L_{H^2} = -\frac{1}{2} \int_{\bf k}\left[{\bf k}^2 H_{{\bf
k}_{\,\mu\nu}} H_{-{\bf k}}^{\mu\nu}-{{\bf k}^2\over 2} H_{\bf k}
H_{-{\bf k}}\right] -\frac{1}{2} \int_{\bf k}\left(
 -\partial_0 H_{{\bf k}_{\,\mu\nu}} \partial^0 H_{-{\bf k}}^{\mu\nu} +
\frac{1}{2} \partial_0 H_{\bf k}  \partial^0 H_{-\bf k}\right),
\end{eqnarray}
where $H_{\bf k}\equiv H^\mu_{{\bf k}\,\mu}$, and the associated
action is $S_{H^2}=\int dx^0 L_{H^2}$.  The terms in the last curved
brackets are suppressed relative to the square brackets terms by a
power of $v^2$. They are treated perturbatively, as operator
insertions, in correlation functions (see discussion below). So,
taking by now only the leading contribution of (\ref{LpotAUX}), it
can be written as
\begin{eqnarray}
\label{Lend:H}\hspace{-0.4cm} L_{H^2} = -\frac{1}{2} \int_{\bf k}
 H_{{\bf -k}_{\,\alpha\beta}} \mathcal{D}_H^{\alpha\beta\mu\nu}
 H_{{\bf k}_{\,\mu\nu}}\,,
\qquad {\rm with}\quad
 \mathcal{D}_H^{\alpha\beta\mu\nu}=
 \frac{1}{2}\left( \eta^{\alpha\mu}\eta^{\beta\nu}
 +\eta^{\alpha\nu}\eta^{\beta\mu}
 -\eta^{\alpha\beta}\eta^{\mu\nu}\right){\bf k}^2\!.
\end{eqnarray}
To find the potential graviton propagator we have to get the
symmetric inverse of the bilinear operator
$\mathcal{D}_H^{\alpha\beta\mu\nu}$. The inverse of its tensorial
part is  $P_{\alpha\beta\mu\nu}$, and the inverse of its momentum
part is simply ${\bf k}^{-2}$. The Feynman rule for the potential
graviton propagator is then
\begin{eqnarray}
\label{FR:PgravProp}
 \vtop{  \vskip-10pt \hbox{\includegraphics[width=3cm]{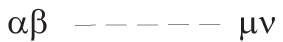} } }
  \:\: &=& \:\:
 \langle H_{{\bf k}_{\,\alpha\beta}} (x^0)
 H_{{\bf q}_{\,\mu\nu}}(x^{\backprime \,0})\rangle \nonumber\\
 &=& -(2\pi)^{d-1}\delta({\bf k} + {\bf
q}){i\over {\bf k}^2}\delta(x^0-x^{\backprime \,0})
P_{\alpha\beta\mu\nu}\,.
\end{eqnarray}
with $P_{\alpha\beta\mu\nu}$ defined in (\ref{FR:RgravPropAUX}). As
shown in section \ref{sec:pc}, the power counting scale for this
propagator is $\langle H_{{\bf k}_{\,\alpha\beta}}
 H_{{\bf q}_{\,\mu\nu}}\rangle \sim r^d v$.

\vskip 0.3cm

To compute diagram \ref{FigFD:EIH}e in the main body of the text
we need the next-to-leading order correction to the potential
graviton propagator. This accounts for the contribution of the
curved brackets terms in (\ref{LpotAUX}) that we neglected to get
the leading propagator (\ref{FR:PgravProp}). The quickest way to
get this propagator correction is to note that the graviton
propagator is proportional to ${1\over k^2}={1\over k_0^2- {\bf
k}^2}$. For a potential graviton, and in the small $v$ limit, one
has ${k_0^2 \over {\bf k}^2}=v^2\ll 1$ which allows the Taylor
expansion,
\begin{equation}
{1\over k_0^2- {\bf k}^2}=-{1\over  {\bf k}^2}\lp 1+{k_0^2 \over
{\bf k}^2} + \mathcal{O}(v^4)\rp .
\end{equation}
Since $k^0 \leftrightarrow \partial_0$ the potential graviton
propagator correction contributing as $r^d v^3$ corresponds to
insert the operator $\partial_0^2$ (which we denote with the
subscript $\otimes$) in the correlation function
(\ref{FR:PgravProp}), yielding
\begin{eqnarray}
\label{FR:PgravPropCorr}
 \vtop{  \vskip-11pt \hbox{\includegraphics[width=3cm]{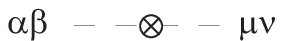} } } \:\: &=& \:\:
 \langle H_{{\bf k}_{\,\alpha\beta}} (x^0)
 H_{{\bf q}_{\,\mu\nu}}(x^{\backprime \,0})\rangle_{\otimes}\nonumber\\
 &=& -(2\pi)^{d-1}\delta({\bf k} + {\bf
q}){i \,\partial_0^2 \over {\bf k}^2}\,\delta(x^0-x^{\backprime
\,0}) P_{\alpha\beta\mu\nu}\,.
\end{eqnarray}
%

\subsection{3-radiation graviton vertex}

To get the 3-radiation graviton vertex using the background field
method we have to go back to (\ref{LgfLghost}) and expand the
background field $g_{(0)}^{\mu\nu}$ up to first order,
\begin{eqnarray}
\label{metricpert:3grav}  g^{(0)}_{\mu\nu}&=& \eta_{\mu\nu}+\kappa_g
\bar{h}_{\mu\nu}\,,
\end{eqnarray}
where the background metric $\eta_{\mu\nu}$ is taken to be
Minkowski spacetime and $\bar{h}_{\mu\nu}$ represent the new small
perturbations around it. Of course, (\ref{metricpert:3grav}) also
induces perturbations on the metric determinant $\sqrt{g_{(0)}}$,
affine connections, Ricci tensor $R^{(0)}_{\mu\nu}$ and Ricci
scalar $R^{(0)}$ that appear in (\ref{metricpert:3grav}).

The result of this expansion is the Lagrangian for the 3-graviton
interaction,
\begin{eqnarray}
\label{L:3grav}
 S_{\bar{h}h^2} = \int d^dx L_{\bar{h}h^2}\,, \qquad
 L_{\bar{h}h^2}(x)
 =-\frac{\kappa_g}{2}\,\bar{h}^{\mu\nu}(x)T^{h^2}_{\mu\nu}(x)\,,
\end{eqnarray}
with
\begin{eqnarray}
\label{Tuv:Prad}
 T^{h^2}_{\mu\nu}(x) \!\!\!&=&\!\!\! -h^{\alpha\beta}\partial_\mu\partial_\nu h_{\alpha\beta}
 +\frac{1}{2}\,h \partial_\mu\partial_\nu h +\left[ \frac{1}{4}\partial_\mu\partial_\nu
 -\frac{3}{8}\,\eta_{\mu\nu} \partial^2\right] \left( h^2-2h^{\alpha\beta}h_{\alpha\beta}\right)
 -\partial^2 \left( h_{\alpha\mu}h^{\alpha}_{\:\:\nu}-h
 h_{\mu\nu}\right)\nonumber\\
 && \hspace{-0.6cm} -\lpp\partial_\alpha\partial_\mu\lp hh_{\nu}^{\:\:\alpha}\rp
  +\partial_\alpha\partial_\nu\lp h h_{\mu}^{\:\:\alpha}\rp\rpp
  +2 \partial_\alpha\partial_\beta\lp h^{\alpha}_{\:\:\mu}h^{\beta}_{\:\:\nu}
  -h^{\alpha\beta}h_{\mu\nu}-\frac{1}{2}\,\eta_{\mu\nu}h^{\alpha\gamma}h^{\beta}_{\:\:\gamma}
  +\frac{1}{2}\,\eta_{\mu\nu}h\,h^{\alpha\beta}  \rp\nonumber\\
 && \hspace{-0.6cm} +2\partial_\alpha\lpp
 h^{\alpha\beta}\lp\partial_\mu h_{\beta\nu}+\partial_\nu h_{\beta\mu} \rp\rpp
 -h^{\alpha}_{\:\:\mu}\partial^2 h_{\alpha\nu}-h^{\alpha}_{\:\:\nu}\partial^2 h_{\alpha\mu}
 +h_{\mu\nu}\partial^2 h \nonumber\\
 && \hspace{-0.6cm}
 +\frac{1}{2}\,\eta_{\mu\nu}\lp h^{\alpha\beta}\partial^2 h_{\alpha\beta}
 -\frac{1}{2}h\partial^2 h  \rp ,
\end{eqnarray}
The Feynman rule for the 3-radiation graviton vertex follows from
its definition in momentum space,
\begin{eqnarray}
\label{3G:VertexDef}
 && \lp V_{\bar{h}h^2}\rp^{\mu\nu}_{\alpha\beta\gamma\sigma}=
i(2\pi)^d \int d^dx \int d^dx_1 d^dx_2 d^dx_3\,
 e^{i(k x_1+p x_2+q x_3)} \frac{\delta}{\delta h^{\alpha\beta}(x_1)}
 \,\frac{\delta}{\delta h^{\gamma\sigma}(x_2)} \,
 \frac{\delta L_{\bar{h}h^2}(x)}{\delta \bar{h}_{\mu\nu}(x_3)}
\,,
 \nonumber\\
&&{\rm with}\qquad \frac{\delta h_{\alpha\beta}(x)}
 {\delta
 h^{\mu\nu}(x_i)}=\delta^d(x-x_i)I_{\alpha\beta\mu\nu}\,,\qquad
 k+p+q=0\,,
\end{eqnarray}
and $I_{\alpha\beta\mu\nu}$ being the symmetric identity tensor,
\begin{eqnarray}
\label{IdDef} I_{\alpha\beta\mu\nu}\equiv \frac{1}{2}
 \left(\eta_{\alpha\mu}\eta_{\beta\nu}+\eta_{\alpha\nu}\eta_{\beta\mu}\right)\,.
\end{eqnarray}
We assumed that all momenta $k,p,q$ is incoming and thus
conservation of momentum in this convention reads as displayed in
the end of (\ref{3G:VertexDef}). After a long but straightforward
computation that also makes use of the integral representation
(\ref{IntRepdelta}) for the delta function we arrive at the
desired Feynman rule for the 3-radiation graviton vertex,
\begin{eqnarray}
\label{FR:3grav} \lp
V_{\bar{h}h^2}\rp^{\mu\nu}_{\alpha\beta\gamma\sigma} \!\!&=&\!\!
 -{i \kappa_g \over 2} \left\{ \left[ k^{\mu} k^{\nu} + (k + q)^{\mu} (k + q)^{\nu} + q^{\mu} q^{\nu} - {3 \over 2} \eta^{\mu
\nu} q^2 \right] \left(I_{\alpha \beta \gamma
\sigma}-\frac{1}{2}\,\eta_{\alpha\beta}\eta_{\gamma\sigma}\right) \right. \nonumber \\
& & + 2q_{\lambda} q_{\delta} \left[ I_{~~~\alpha \beta}^{\lambda
\delta} I_{~~~\gamma \sigma}^{\mu \nu} + I_{~~~\gamma
\sigma}^{\lambda \delta} I_{~~~\alpha \beta}^{\mu \nu} -
I_{~~~\alpha \beta}^{\mu\delta} I_{~~~\gamma \sigma}^{\nu\lambda} -
I_{~~~\alpha \beta}^{ \nu\lambda}
I_{~~~\gamma \sigma}^{ \mu\delta} \right] \nonumber \\
& &+  q_{\lambda} q^{\mu} \left( \eta_{\alpha \beta} I_{~~~\gamma
\sigma}^{\lambda \nu} + \eta_{\gamma \sigma} I_{~~~\alpha
\beta}^{\lambda \nu} \right) + q_{\lambda} q^{\nu} \left(
\eta_{\alpha \beta} I_{~~~\gamma \sigma}^{\lambda \mu} +
\eta_{\gamma \sigma} I_{~~~\alpha
\beta}^{\lambda \mu} \right)  \nonumber \\
& & - q^2 \left( \eta_{\alpha \beta} I_{~~~\gamma \sigma}^{\mu \nu}
- \eta_{\gamma \sigma} I_{~~~\alpha \beta}^{\mu \nu} \right) -
\eta^{\mu \nu} q^{\lambda} q^{\delta} \left( \eta_{\alpha \beta}
I_{\gamma \sigma \delta\lambda} + \eta_{\gamma \sigma} I_{\alpha
\beta \delta\lambda}
\right)  \nonumber \\
& & -  2 q^{\lambda} \left[  I_{ \alpha \beta\lambda \delta}\lp
I_{~~~\gamma \sigma}^{\delta \nu} (k + q)^{\mu}
 +  I_{~~~\gamma \sigma}^{\delta \mu}  (k +
 q)^{\nu}\rp
  - I_{\gamma \sigma \lambda \delta} \lp I_{~~~\alpha\beta}^{\delta \nu}  k^{\mu} + I_{~~~\alpha\beta}^{\delta \mu}
 k^{\nu} \rp\right]
\nonumber \\
& &  + q^2 \left( I_{~~~\alpha \beta}^{\delta \mu} I_{\gamma \sigma
\delta}^{~~~~\nu} + I_{\alpha \beta \delta}^{~~~~\nu} I_{~~~\gamma
\sigma}^{\delta \mu} \right) + \eta^{\mu \nu} q^{\lambda} q_{\delta}
\left( I_{\alpha \beta \lambda \rho} I_{~~\gamma \sigma}^{\rho
\delta} + I_{\gamma \sigma \lambda \rho} I_{~~~\alpha \beta}^{\rho
\delta} \right)
\nonumber \\
& & +  \left( k^2 + (k+q)^2 \right) \left[ I_{~~~\alpha
\beta}^{\delta \mu} I_{\gamma \sigma\delta}^{~~~~\nu} + I_{~~~\alpha
\beta}^{\delta \nu} I_{\gamma \sigma\delta }^{~~~~\mu} - {1 \over 2}
\eta^{\mu \nu} \left(I_{\alpha \beta \gamma
\sigma}-\frac{1}{2}\,\eta_{\alpha\beta}\eta_{\gamma\sigma}\right) \right]  \nonumber \\
& &  \left. - \left( k^2 \eta_{\alpha \beta} I_{~~~\gamma
\sigma}^{\mu \nu} + (k + q)^2 \eta_{\gamma \sigma} I_{~~\alpha
\beta}^{ \mu \nu} \right)  \right\}\,.
\end{eqnarray}
This Feynman rule is independent of the dimensionality of the
spacetime (contrary to the rule for the graviton propagator and
potential graviton vertices). In particular, (\ref{FR:3grav}) is
the same as the Feynman rule first obtained in 4-dimensions by
\cite{Donoghue:1994dn,BjerrumBohr:2002kt} (for the details see
\cite{Bohr:2001}). The 3-point correlation function for the
3-radiation graviton interaction, $\langle
h_{\mu\nu}h_{\alpha\beta}h_{\gamma\sigma} \rangle$, can be
obtained trivially from the tensorial product of this vertex rule
with the radiation graviton propagators (\ref{FR:RgravProp}),
\vskip0.7cm
\begin{eqnarray}
\label{FR:3grav:Rad}
  \vtop{  \vskip-33pt  \hbox{ \includegraphics[width=3cm]{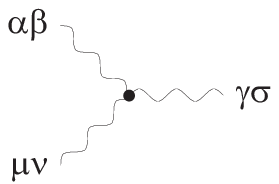} } }
 &=&
\langle h_{\alpha\beta}(x_1)h_{\mu\nu}(x_2)h_{\gamma\sigma}(x_3)
\rangle
 \\
 &=& \langle  h_{\alpha\beta}(x_1)
 h_{\hat{\alpha}\hat{\beta}}(x^{\backprime}_1)\rangle
\lp V_{\bar{h}H^2}\rp^{\hat{\alpha}\hat{\beta}
        \hat{\mu}\hat{\nu}\hat{\gamma}\hat{\sigma}}
 \langle h_{\mu\nu} (x_2)
 h_{\hat{\mu}\hat{\nu}}(x^{\backprime}_2)\rangle
 \langle h_{\gamma\sigma}(x_3)
 h_{\hat{\gamma}\hat{\sigma}}(x^{\backprime}_3)\rangle\,.\nonumber
\end{eqnarray}
We do not need (\ref{FR:3grav}) or (\ref{FR:3grav:Rad}) (we just
quote them for completeness), but we will make use of
(\ref{Tuv:Prad}) in the next subsections.

\subsection{3--radiation-potential graviton vertex and correlation function\label{sec:FR:vertexPotRad}}

The action describing the 3--radiation-potential graviton
interaction is
\begin{eqnarray}
\label{L:3grav:PotRad}
 S_{\bar{h}H^2} =\int d^dx L_{\bar{h}H^2}\,, \qquad
 L_{\bar{h}H^2}(x)
 =-\frac{\kappa_g}{2}\,\bar{h}^{\mu\nu}(x)T^{H^2}_{\mu\nu}(x)\,,
\end{eqnarray}
where $T^{H^2}_{\mu\nu}(x)$ is obtained from $T^{h^2}_{\mu\nu}(x)$
by doing the replacement $h\rightarrow H$ in (\ref{Tuv:Prad}).

We proceed using the Fourier transform (\ref{FourierHk}) and keeping
only the leading order terms in the velocity. This yields
\begin{eqnarray}
\label{Tuv:Pgrav}
  &&\hspace{-0.5cm} T^{H^2}_{\mu\nu}(x) =\int_{\bf k}
\int_{\bf q}
 e^{i({\bf k}+{\bf q})\cdot {\bf x}} \,\mathcal{T}^{H^2}_{\mu\nu}(x^0)\,,\nonumber\\
  && \mathcal{T}^{H^2}_{\mu\nu} (x^0)= \lp
{\bf q}_u {\bf q}_v+\frac{1}{2}\,\eta_{\mu\nu}\,{\bf q}^2 \rp \lp
H_{\bf k}^{\alpha\beta}
 H_{{\bf q}_{\,\alpha\beta}} -\frac{1}{2}\,H_{\bf k}H_{\bf q}\rp
    -{\bf q}^2 \lp H_{{\bf k}_{\,\mu}}^{\alpha}
 H_{{\bf q}_{\,\alpha\nu}} + H_{{\bf k}_{\,\nu}}^{\alpha}
 H_{{\bf q}_{\,\alpha\mu}} - H_{{\bf k}_{\,\mu\nu}}H_{\bf q} \rp\nonumber\\
 && \hspace{1.2cm}
 +\frac{1}{2}\lp {\bf k}_u {\bf k}_v+{\bf q}_u {\bf q}_v+ 2{\bf k}_{(u} {\bf q}_{\,v)}
  +\frac{3}{2}\eta_{\mu\nu}\lp {\bf k}^2+{\bf q}^2+2{\bf
k}\cdot{\bf q}\rp \rp \lp H_{\bf k}^{\alpha\beta}
 H_{{\bf q}_{\,\alpha\beta}} -\frac{1}{2}\,H_{\bf k}H_{\bf q}\rp \nonumber\\
 && \hspace{1.2cm}
 +\lp {\bf k}_i {\bf k}_u+{\bf q}_i {\bf q}_u+ 2{\bf k}_{(i} {\bf q}_{\,u)} \rp H_{\bf k} H_{{\bf q}_{\,\nu}}^{i}
 +\lp {\bf k}_i {\bf k}_v+{\bf q}_i {\bf q}_v+ 2{\bf k}_{(i} {\bf q}_{\,v)} \rp H_{\bf k} H_{{\bf q}_{\,\mu}}^{i}\nonumber\\
 && \hspace{1.2cm}
 -2\lp {\bf k}_i {\bf k}_j+{\bf q}_i {\bf q}_j+2{\bf k}_{(i} {\bf q}_{\,j)} \rp \lp  H_{{\bf k}_{\,\mu}}^{i} H_{{\bf q}_{\,\nu}}^{j}
 -H_{\bf k}^{ij} H_{{\bf q}_{\,\mu\nu}}-\frac{1}{2}\,\eta_{\mu\nu}
 \lp H_{\bf k}^{i\alpha}H_{{\bf q}_{\,\alpha}}^{j}-H_{\bf k}H_{\bf q}^{ij} \rp \rp\nonumber\\
 && \hspace{1.2cm}
 -2\lp {\bf q}_i {\bf q}_u+{\bf k}_i {\bf q}_u\rp H_{\bf k}^{i\alpha} H_{{\bf q}_{\,\alpha\nu}}
 -2\lp {\bf q}_i {\bf q}_v+{\bf k}_i {\bf q}_v\rp H_{\bf k}^{i\alpha} H_{{\bf q}_{\,\alpha\mu}}\nonumber\\
 && \hspace{1.2cm}
 -\lp {\bf k}^2+{\bf q}^2+2{\bf k}\cdot{\bf q}\rp
   \lp H_{{\bf k}_{\,\alpha\mu}} H_{{\bf q}_{\,\nu}}^{\alpha}-H_{\bf k}H_{{\bf q}_{\,\mu\nu}} \rp
 + {\rm higher}\:\,{\rm order}\:\,v\:\,{\rm terms}\label{CalTuv:Pgrav}.
\end{eqnarray}
Notice that in this expression we use the Latin letters $u,v$ when
only spatial $\mu,\nu$ make a leading order contribution, and we
use parenthesis () around the indices to denote symmetrization,
 ${\bf k}_{(i} {\bf q}_{\,j)}=\lp {\bf k}_i {\bf q}_j+{\bf q}_i {\bf k}_j\rp /2$.

Proceeding, we do the multipole expansion (\ref{MultExp}) of ${\bar
h}_{\mu\nu}(x)$ around the system's center of mass ${\bf X}$, and
keep only the leading order term in the velocity $v$, ${\bar
h}_{\mu\nu}(x^0,{\bf X})$.
 So, to leading order ${\bar h}_{\mu\nu}(x)$ only depends on
$x^0$ but not on ${\bf x}$. This allows to write
(\ref{L:3grav:PotRad}) in lowest order as
\begin{eqnarray}
\label{L:3grav:PotRad2}
 S_{\bar{h}H^2} =\int dx^0 {\bar h}_{\mu\nu}(x^0,{\bf X})
\int_{\bf k} \int_{\bf q} \int d^{d-1}{\bf x}\,
 e^{i({\bf k}+{\bf q})\cdot {\bf x}}
 \lp -\frac{\kappa_g}{2}\,\mathcal{T}^{H^2}_{\mu\nu}(x^0)\rp\,.
\end{eqnarray}
The spatial integral can be done using the integral representation
of the delta function $\int_{\bf x}\,e^{i({\bf k}+{\bf
q})\cdot{\bf x}}=\delta{({\bf k}+{\bf q})}$. This operation
results in ${\bf q}\rightarrow -{\bf k}$ in (\ref{Tuv:Pgrav}) and
only the first line survives to the procedure. We can finally
write the leading order action for the 3-graviton interaction
between potential and radiation gravitons as
\begin{eqnarray}
 \label{L:3grav:PotRadEnd}
 S_{\bar{h}H^2} \!\!&=&\!\! \int dx^0
  L_{\bar{h}H^2}(x^0,{\bf X})\,, \qquad {\rm with}
   \nonumber\\
 L_{{\bar h} H^2}(x^0,{\bf X}) \!\!&=&\!\! \kappa_g
 {\bar h}^{00} \int_{\bf k} {\bf k}^2 \left[
 H_{{\bf k}_0}^\mu H_{-{\bf k}_{\,\mu 0}}
 - {1\over 2} H_{{\bf k}_{\,00}} H_{\bf -k}
 -{1\over 4} H_{\bf k}^{\mu\nu}
H_{-{\bf k}_{\,\mu\nu}} + {1\over 8} H_{\bf k}  H_{-{\bf k}} \right]
 \nonumber\\
 &&
 + \kappa_g{\bar h}^{0i} \int_{\bf k} {\bf k}^2
 \left[2  H^{\mu}_{{\bf k}_{\,0}} H_{{-\bf k}_{\,\mu i}}
- H_{{\bf k }_{\,0i}} H_{-\bf k}\right]\nonumber\\
 &&
  + \kappa_g{\bar h}^{ij}\int_{\bf k}  {\biggl [}
  -{1\over 2} {\bf k}_i {\bf k}_j\lp
H_{\bf k}^{\mu\nu} H_{{-{\bf k}}_{\,\mu\nu}} - {1\over 2} H_{\bf k}
H_{-{\bf k}} \rp
 + {\bf k}^2
 H_{{\bf k}_{\,i \mu}} H_{-{\bf k}_{\,j}}^\mu   \nonumber\\
 &&  \qquad\qquad \quad- {1\over 2}{\bf k}^2
  H_{{\bf k}_{\,ij}} H_{\bf -k} -\frac{{\bf k}^2}{4}\,\eta_{ij} \left(
  H_{\bf k}^{\mu\nu} H_{-{\bf k}_{\,\mu\nu}} - {1\over 2} H_{\bf k}
   H_{-{\bf k}}\right) {\biggr ]},
\end{eqnarray}
where ${\bar h}^{00},{\bar h}^{0i},{\bar h}^{ij}$ are functions of
$(x^0,{\bf X})$ and $H_{\bf k}^{\mu\nu}=H_{\bf k}^{\mu\nu}(x^0)$.

The Feynman rule for the 3--radiation-potential graviton vertex
follows from the definition,
\begin{eqnarray}
\label{3G:VertexDef:RadPot}
 && \lp V_{\bar{h}H^2}\rp_{\alpha\beta\mu\nu}=
i\int dx^0 \int dx^0_1 dx^0_2\, \frac{\delta}{\delta
 H_{\bf k_1}^{\alpha\beta}(x^0_1)} \,
 \frac{\delta L_{{\bar h} H^2}(x^0,{\bf X})} {\delta H_{\bf k_2}^{\mu\nu}(x^0_2)}
\,,
\end{eqnarray}
and use of the functional derivative,
\begin{eqnarray}
\label{Def:derivPot}
 \frac{\delta H_{{\bf k}_{\,\alpha\beta}}(x^0)}
 {\delta H_{\bf q}^{\mu\nu}(x^0_i)}=\delta\lp{\bf k}+{\bf q}\rp\,\delta\lp x^0-x^0_i\rp\, I_{\alpha\beta\mu\nu}\,,
\end{eqnarray}
with $I_{\alpha\beta\mu\nu}$ defined in (\ref{IdDef}). Finally,
the 3-point correlation function for the interaction between two
potential gravitons and radiation graviton is (dropping the bars
over the graviton)
\vskip0.7cm
\begin{eqnarray}
\label{FR:3grav:RadPot}
 &&\hspace{-1cm}
  \vtop{  \vskip-40pt  \hbox{ \includegraphics[width=3cm]{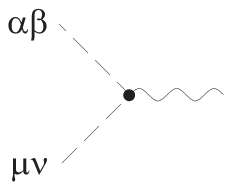} } }
 \:\:=\:\:
 \langle h(x^0,X) H_{{\bf k}_{\,\alpha\beta}}(x^0_1) H_{{\bf q}_{\,\mu\nu}} (x^0_2)\rangle
 \\
 && \quad = \langle H_{{\bf k}_{\,\alpha\beta}}(x^0_1)
 H_{{\bf k^{\backprime}}_{\,\gamma\sigma}}(x^{\backprime \,0}_1)\rangle
\lp V_{\bar{h}H^2}\rp^{\gamma\sigma\lambda\eta}
 \langle H_{{\bf q}_{\,\mu\nu}} (x^0_2)
 H_{{\bf q^{\backprime}}_{\,\lambda\eta}}(x^{\backprime \,0}_2)\rangle\,\nonumber\\
 && \quad = -i\kappa_g (2\pi)^{d-1}\delta(x^0_1-x^0_2)
 \delta({\bf k} + {\bf q}) \frac{1}{{\bf k}^2{\bf q}^2}\nonumber\\
&& \qquad \times {\biggl\{ } h^{00}  {\bf k}^2 \left[
 - {1\over 2}\,P_{\alpha\beta\mu\nu}+P_{\alpha\beta\mu 0}\,\eta_{0\nu}+P_{\alpha\beta 0\nu}\,\eta_{0\mu}
 -\frac{2}{d-2}\lp P_{\alpha\beta 00}\,\eta_{\mu\nu}+\eta_{\alpha\beta}\,P_{00\mu\nu}\rp
 \right]\nonumber\\
&& \qquad\qquad +h^{0i}  {\bf k}^2 \left[ 2\lp P_{\alpha\beta\mu
i}\,\eta_{0\nu}+P_{\alpha\beta i\nu}\,\eta_{0\mu} \rp
 +\frac{4}{d-2}\lp \eta_{\alpha\beta}\,P_{0i\mu\nu}-P_{\alpha\beta 0i}\,\eta_{\mu\nu}\rp
 \right]\nonumber\\
&& \qquad\qquad +h^{ij}  \left[  -\lp {\bf k}_i{\bf k}_j + {1\over
2}\eta_{ij}\, {\bf k}^2 \rp P_{\alpha\beta\mu\nu}+
 {\bf k}^2 \lp
I_{\alpha\beta\mu j}\,\eta_{i\nu}+I_{\alpha\beta j\nu}\,\eta_{i\mu}
 -\frac{2}{d-2}\, I_{\alpha\beta ij}\,\eta_{\mu\nu}\rp
 \right] {\biggr\} }.\nonumber
\end{eqnarray}
In the main text we need only the $\alpha,\beta,\mu,\nu=0$ component
which reads
\begin{eqnarray}
\label{FR:3grav:RadPot2}
 &&\hspace{-1cm}\langle h(x^0,X) H_{{\bf k}_{\,00}}(x^0_1) H_{{\bf q}_{\,00}} (x^0_2)\rangle
 \\
 && \quad = -i\kappa_g (2\pi)^{d-1}\delta(x^0_1-x^0_2)
 \delta({\bf k} + {\bf q}) \frac{1}{{\bf k}^2{\bf q}^2}\, \frac{d-3}{d-2}
 \lpp {3\over 2}\,{\bf k}^2 h^{00}- \lp {1\over 2}\,{\bf k}^2 \eta_{ij} +{\bf k}_i{\bf k}_j \rp h^{ij}
 \rpp .
\nonumber
\end{eqnarray}
Using the power counting rules of Table \ref{PCtable} we find that
 $\langle h H_{{\bf k}_{\,\alpha\beta}} H_{{\bf q}_{\,\mu\nu}}\rangle \sim L^{-1/2}
v^{d+3 \over 2}r^{d}$.

\subsection{3-potential graviton vertex and correlation function\label{sec:3PotVertex}}

To obtain the Lagrangian for the 3-potential graviton interaction
we begin by making the replacement $\bar{h}^{\mu\nu}\rightarrow
\bar{H}^{\mu\nu}$ in (\ref{L:3grav:PotRad}). We then take the
Fourier transform (\ref{FourierHk}) of the potential graviton
$\bar{H}^{\mu\nu}$ and use the integral representation of the
delta function (\ref{IntRepdelta}) to get
\begin{eqnarray}
\label{L:3grav:Pot}
 && S_{\bar{H}H^2} =\int dx^0 L_{\bar{H} H^2}\,,
\qquad
 L_{\bar{H}H^2}(x^0)
 =-\frac{\kappa_g}{2}\int_{\bf k}\int_{\bf q}\int_{\bf p} (2\pi)^{d-1}
 \delta\lp {\bf k}+{\bf q}+{\bf p} \rp \bar{H}_{\bf
 p}^{\mu\nu}(x^0)\,\mathcal{T}^{H^2}_{\mu\nu}(x^0)\,,\nonumber\\
&&
\end{eqnarray}
where $\mathcal{T}^{H^2}_{\mu\nu}(x^0)$ is defined in
(\ref{CalTuv:Pgrav}).

The Feynman rule for the 3-potential graviton vertex follows from
the definition,
\begin{eqnarray}
\label{3G:VertexDef:pot}
 && \lp V_{\bar{H}H^2}\rp_{\alpha\beta\mu\nu\gamma\sigma}=
i\int dx^0 \int dx^0_1 dx^0_2 dx^0_3\, \frac{\delta}{\delta
 H_{\bf k_1}^{\alpha\beta}(x^0_1)} \, \frac{\delta}{\delta
 H_{\bf k_2}^{\mu\nu}(x^0_2)} \,
 \frac{\delta L_{{\bar H} H^2}(x^0)} {\delta H_{\bf k_3}^{\gamma\sigma}(x^0_3)}
\,,
\end{eqnarray}
and use of (\ref{Def:derivPot}). The 3-point correlation function
for the interaction between three potential gravitons is finally
\vskip0.7cm
\begin{eqnarray}
\label{FR:3grav:Pot}
 &&\hspace{-1cm}
  \vtop{  \vskip-35pt  \hbox{ \includegraphics[width=3cm]{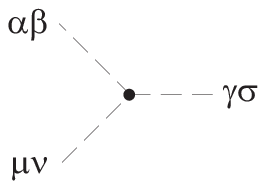} } }
  \:\: = \:\:
 \langle H_{{\bf k}_{\,\alpha\beta}}(x^0_1) H_{{\bf q}_{\,\mu\nu}} (x^0_2)
  H_{{\bf p}_{\,\gamma\sigma}} (x^0_3)\rangle
 \\
 && \quad = \langle H_{{\bf k}_{\,\alpha\beta}}(x^0_1)
 H_{{\bf k^{\backprime}}_{\,\hat{\alpha}\hat{\beta}}}(x^{\backprime \,0}_1)\rangle
\lp V_{\bar{H}H^2}\rp^{\hat{\alpha}\hat{\beta}
        \hat{\mu}\hat{\nu}\hat{\gamma}\hat{\sigma}}
 \langle H_{{\bf q}_{\,\mu\nu}} (x^0_2)
 H_{{\bf q^{\backprime}}_{\,\hat{\mu}\hat{\nu}}}(x^{\backprime \,0}_2)\rangle
 \langle H_{{\bf p}_{\,\gamma\sigma}}(x^0_3)
 H_{{\bf p^{\backprime}}_{\,\hat{\gamma}\hat{\sigma}}}(x^{\backprime \,0}_3)\rangle\,\nonumber\\
 &&
 \quad = -\frac{\kappa_g}{2} (2\pi)^{d-1}\delta(x^0_1-x^0_2)\delta(x^0_1-x^0_3)
 \delta\lp {\bf k} + {\bf q}+ {\bf p}\rp \frac{1}{{\bf k}^2\,{\bf q}^2\,{\bf p}^2}\nonumber\\
&& \qquad \times {\biggl\{ }
 2P_{\alpha\beta\mu\nu}P^{ij}_{~~\gamma\sigma}
 \lp {\bf k}_i {\bf k}_j+{\bf q}_i {\bf q}_j +{\bf k}_{(i} {\bf q}_{\,j)} \rp\nonumber\\
&& \qquad \quad
 -\lpp 2P^{\lambda\eta}_{~~~\gamma\sigma}
    \lp P_{\alpha\beta\eta(\mu}\,\eta_{\,\nu)\lambda}+\frac{1}{d-2}\,P_{\lambda\eta\alpha\beta}\,\eta_{\mu\nu} \rp
    +\frac{3}{d-2}\,P_{\alpha\beta\mu\nu}\,\eta_{\gamma\sigma}\rpp\lp{\bf k}^2+{\bf q}^2+2{\bf k}\cdot{\bf q}\rp
\nonumber\\
&& \qquad\quad
 +\lpp
P_{\gamma\sigma\alpha(\mu}\eta_{\,\nu)\beta}+P_{\gamma\sigma\beta(\mu}\eta_{\,\nu)\alpha}
 +\frac{1}{d-2}\,\lp 2P_{\alpha\beta\gamma\sigma}\,\eta_{\mu\nu}
 -P_{\alpha\beta\mu\nu}\,\eta_{\gamma\sigma}\rp \rpp\lp{\bf k}^2+{\bf q}^2\rp  \nonumber\\
&& \qquad \quad
   -4\lp {\bf k}_i {\bf k}_j+{\bf q}_i {\bf q}_j +2{\bf k}_{(i} {\bf q}_{\,j)} \rp\nonumber\\
&& \qquad \qquad \times {\biggl [}
  P^{i\lambda}_{~~~\gamma\sigma}
   \lp\frac{2}{d-2}\,P^{j}_{~\lambda\alpha\beta}\,\eta_{\mu\nu}+P_{\lambda\eta\alpha\beta}P^{j\eta}_{~~~\mu\nu}\rp
 + P^{\lambda\eta}_{~~~\gamma\sigma}
 \lp P^{i}_{~\lambda\mu\nu}P^{j}_{~\eta\alpha\beta}-P_{\lambda\eta\alpha\beta}P^{ij}_{~~\mu\nu}\rp \nonumber\\
&& \qquad \qquad\quad
 +\frac{1}{d-2}\lp\frac{1}{d-2}\,P^{ij}_{~~\alpha\beta}\,\eta_{\mu\nu}\,\eta_{\gamma\sigma}
  + I_{\alpha\beta(\mu}^{~~~~~j}  \,\eta_{\,~\nu)}^{i}\eta_{\gamma\sigma}
  -\frac{1}{d-2}\eta_{\alpha\beta}\,\eta_{\gamma\sigma}\,I^{ij}_{~~\mu\nu}\rp
 {\biggr ]} {\biggr\} },\nonumber
\end{eqnarray}
where we used (again) the rule that a sum over Latin indices means
that only spatial indices contribute (to leading order), and the
parenthesis () around the indices denote symmetrization. In the
main text we need only the $\alpha,\beta,\mu,\nu,\gamma,\sigma=0$
component which reads
\begin{eqnarray}
\label{FR:3grav:Pot2}
 &&\hspace{-1cm} \langle H_{{\bf k}_{\,00}}(x^0_1) H_{{\bf q}_{\,00}} (x^0_2)
  H_{{\bf p}_{\,00}} (x^0_3)\rangle
 \\
 && \quad = \kappa_g (2\pi)^{d-1}\delta(x^0_1-x^0_2)\delta(x^0_1-x^0_3)
 \delta\lp {\bf k} + {\bf q}+ {\bf p}\rp \frac{(d-3)^{2}}{(d-2)^{2}}\,
 \frac{{\bf k}^2+{\bf q}^2+{\bf p}^2}{{\bf k}^2\,{\bf q}^2\,{\bf p}^2}.
\nonumber
\end{eqnarray}
Using the power counting rules of Table \ref{PCtable} we find that
$\langle H_{{\bf k}_{\,\alpha\beta}} H_{{\bf q}_{\,\mu\nu}}
  H_{{\bf p}_{\,\gamma\sigma}}\rangle\sim L^{-1/2} v^{7/2}r^{3d/2}$.

\subsection{Point-particle vertices}

The Feynman rules for the non-linear vertex interactions of
particles with potential gravitons are obtained by taking the
small velocity expansion (\ref{Spp:exp}) of the point-particle
action $S_{pp}$, with $\delta g_{\mu\nu}\rightarrow H_{\mu\nu}$.
In the expansion (\ref{Spp:exp}) we only display the lowest order
terms that contribute to the leading order observables of
interest. We also take the Fourier transform (\ref{FourierHk}) of
the potential graviton to keep a well-defined track of the
velocity power counting.

The Feynman rules for the interaction vertex between the particle
and a potential graviton vertex then follows from the definition of
vertex interaction factor,
\begin{eqnarray}
\label{VertexDef} V_{\mu\nu}=i\,\frac{\delta S_{pp}}
 {\delta H_{\bf k}^{\mu\nu}}\,,\qquad V_{\mu\nu\alpha\beta}=i\,\frac{\delta^2 S_{pp}}
 {\delta H_{\bf k}^{\mu\nu} \delta H_{\bf q}^{\alpha\beta}}\,,
\end{eqnarray}
and use of (\ref{Def:derivPot}).
 The point-particle vertex Feynman rules are then (following the order
 of appearance in (\ref{Spp:exp}))
\begin{eqnarray}
\label{FR:VertexRules}
 && V_{\mu\nu}^{(1)}=-\frac{i\kappa_g m_a}{2}\int
dx_a^0\int_{\bf k}e^{i{\bf k}\cdot {\bf
x}}\,\eta_{0\mu}\eta_{0\nu}\,,  \nonumber\\
&& V_{\mu\nu}^{(2)}=-\frac{i\kappa_g m_a}{2} \int dx_a^0\int_{\bf
k}e^{i{\bf k}\cdot {\bf x}}\,
v_a^i\left(\eta_{0\mu}\eta_{i\nu}+\eta_{0\nu}\eta_{i\mu}\right)\,,   \nonumber\\
&& V_{\mu\nu}^{(3)}=-\frac{i\kappa_g m_a}{2} \int dx_a^0\int_{\bf
k}e^{i{\bf k}\cdot {\bf x}}\,
v_a^i v_a^j\left(\eta_{i\mu}\eta_{j\nu}+\eta_{i\nu}\eta_{j\mu}\right)\,, \nonumber\\
&& V_{\mu\nu}^{(4)}=-\frac{i\kappa_g m_a}{4} \int dx_a^0\int_{\bf
k}e^{i{\bf k}\cdot {\bf x}} \,{\bf v}^2_a\,
\eta_{0\mu}\eta_{0\nu}\,, \nonumber\\
&& V_{\mu\nu\alpha\beta}^{(5)}=\frac{i\kappa_g^2 m_a}{4} \int
dx_a^0\int_{\bf k}\int_{\bf q}e^{i({\bf k}+{\bf q})\cdot {\bf x}}
 \,\eta_{0\mu}\eta_{0\nu}\eta_{0\alpha}\eta_{0\beta}\,.
\end{eqnarray}
The power counting rules for the vertices are shown in Table
\ref{PCtable2} (the kinetic contributions in (\ref{Spp:exp}), \ie
the two last terms, have power counting rules of $L^{1/2} v^0$ and
$L^{1/2} v^2$, respectively).

\setcounter{equation}{0}
 \section{Useful relations for Feynman diagram computations}
 \label{sec:AUX}

In this appendix we present some standard formulas that are useful
to compute Feynman diagram contributions.

We start with the integral representation of the delta function
\begin{equation}\label{IntRepdelta}
\delta({\bf k})\equiv \delta^{d-1}({\bf k})
 =\int{d^{d-1}{\bf x}\over (2\pi)^{d-1}}e^{i{\bf k}\cdot{\bf x}}\equiv \int_{\bf x} e^{i{\bf k}\cdot{\bf x}}\,,
 \qquad \delta({\bf x})= \int_{\bf k} e^{i{\bf k}\cdot{\bf x}}.
\end{equation}
and the well known integrals,
\begin{equation}\label{deltaInt}
\int f(z)\delta(z-a)=f(a)\,, \qquad \int f'(z)\delta(z-a)=-f'(a)\,,
\end{equation}
where $f'(z)={df\over dz}$.

 The integrals over the momentum are computed using the
following relations,
\begin{eqnarray}\label{IntMomPos}
\int {d^n{\bf k}\over (2\pi)^n}\, {1\over ({\bf k}^2)^\alpha}\,
e^{-i{\bf k}\cdot {\bf x}} = {1\over (4 \pi)^{n/2}}
{\Gamma(n/2-\alpha)\over \Gamma(\alpha)} \left({{\bf x}^2\over
4}\right)^{\alpha-n/2},
\end{eqnarray}
\begin{eqnarray}\label{IntMomPos2}
\int {d^{d-1}{\bf k}\over (2\pi)^{d-1}}\,
 {{\bf k}_i {\bf k}_j \over {\bf k}^4} \, e^{-i{\bf k}\cdot {\bf x}} =
 \frac{\Gamma\lp {d-3\over 2} \rp} {8 \pi^{d-1\over 2}}
|{\bf x}|^{-(d-3)}\lpp \delta_{ij}-(d-3)\,
 \frac{{\bf x}_i {\bf x}_j}{|{\bf x}|^2}\rpp,
\end{eqnarray}
and
\begin{eqnarray}\label{IntMom}
&&\int {d^{2\eta}{\bf k}\over (2\pi)^{2\eta}}
 {1\over {\bf k}^2\lp {\bf k}-{\bf p}\rp^2}=I_0
  \lp{\bf p}^2\rp^{\eta-2},\qquad I_0\equiv
  {\Gamma(2-\eta)\lpp\Gamma(\eta-1)\rpp^2 \over (4\pi)^{\eta}\Gamma(2\eta-2)}\,,\nonumber\\
&&
 \int {d^{2\eta}{\bf k}\over (2\pi)^{2\eta}}
 {{\bf k}_i\over {\bf k}^2\lp {\bf k}-{\bf p}\rp^2}={1\over 2}\,I_0
 {\bf p}_i \lp{\bf p}^2\rp^{\eta-2},\nonumber\\
&&
 \int {d^{2\eta}{\bf k}\over (2\pi)^{2\eta}}
 {{\bf k}_i {\bf k}_j\over {\bf k}^2\lp {\bf k}-{\bf p}\rp^2}
 =-{\delta_{ij}\over 4(2\eta-1)}\,I_0\lp{\bf p}^2\rp^{\eta-1}+
{\bf p}_i {\bf p}_j {\eta\over 2(2\eta-1)}\,I_0\lp{\bf
p}^2\rp^{\eta-2}.
\end{eqnarray}
The integral (\ref{IntMomPos}) is needed to evaluate all Feynman
diagrams. Relation (\ref{IntMomPos2}) is needed to compute
(\ref{FD:Lrad:e}), and relations (\ref{IntMom}) are needed to
compute (\ref{FD:EIH:d}).


\end{document}